\documentclass[extra,referee]{gji}
\usepackage{amsmath}
\usepackage{bm}
\usepackage[hidelinks]{hyperref}
\usepackage{enumitem}
\usepackage{color}
\usepackage{soul}
\usepackage[normalem]{ulem}
\usepackage{graphicx}
\usepackage{lineno} 
\usepackage{amsmath,amssymb}
\usepackage{gensymb}
\usepackage[nodisplayskipstretch]{setspace}
\usepackage{caption}
\usepackage{titlesec}
\titleformat{\subsubsection}[runin]
{\normalfont\bfseries}{\thesubsubsection}{1em}{}
\begin{document}

\begin{center}
\textbf{\Large {Friction Laws and numerical modeling of the seismic cycle}} \\[20pt]
\textcolor{blue}{\small To appear as a chapter in ``The Seismic Cycle: From Observation to Modeling edited by F. Rolandone''}\\[20pt]

Marion Y. Thomas$^1$ and Harsha S. Bhat$^2$

\begin{enumerate}[leftmargin=*,labelindent=0pt,label=\roman*.]
\small
\it
\item{Institut des Sciences de la Terre Paris, Sorbonne Universit\'e, CNRS-UMR 7193, Paris, France.}
\item{Laboratoire de G\'{e}ologie, \'{E}cole Normale Sup\'{e}rieure, CNRS-UMR 8538, PSL Research University, Paris, France.}
\end{enumerate}

\end{center}

\section{Friction Laws}
\label{sec:frictionlaw}
\subsection{Historical notions about friction}
\label{subsec:frichistory}
Friction is resistance to motion that appears when two surfaces in contact slide against one another. Generally speaking, the concept of `friction' describes the dissipation of energy that occurs.
 Most phenomena associated with sliding friction can be understood from observations made by Leonardo da Vinci. He was the first to note that, based on his experiments, friction is proportional to 1/4th of the applied pressure and that it is independent of the area of contact between two active surfaces. This latter observation was inspired by the fact that the resistance to sliding of a coil of rope is the same as for a stretched piece of rope.

Almost two centuries later, in the  $18^\mathrm{th}$ century,  Guillaume Amontons and Charles-Augustin de Coulomb, carried out rigorous experiments on friction, with the aim of obtaining quantitative results. The collective work by L. da Vinci, G. Amontons and C.-A. de Coulomb led to the two fundamental 'laws' of friction. These statements, simple and still valid, are widely applicable:
\begin{itemize}[
  align=left,
  leftmargin=2em,
  itemindent=0pt,
  labelsep=0pt,
  labelwidth=2em
]
\item the friction force acting between two sliding surfaces is proportional to the load pressing the surfaces together. That is, these two forces have a constant ratio, often called the coefficient of friction.
\item the sliding force is independent of the apparent area of contact between the two surfaces.
\end{itemize}

The discoveries that followed (cf. Chapter 1), led researchers to revisit these laboratory experiments in order to better understand earthquakes. In 1966, in a now-famous paper, Brace and Byerlee showed that the creation of new fractures 
was not the only model that could explain the existence of seismic faults \cite{brace1966b}. 
In their experimental protocol, they pre-cut a rock sample and loaded its extremities, while also applying confining pressure. They observed that the sliding between the two pieces of rock was not continuous, but a jerky motion with accelerations and decelerations. This was the origin of the theory, which is widely accepted today, that earthquakes are governed by frictional forces. 

\subsection{From static friction to dynamic friction}
\label{subsec:muSmuD}

If we go back to the fundamental laws of friction stated by Amonton-Coulomb, they are mathematically expressed as follows.  The frictional force $\mathbf{F}_{fric} = \tau A$ is independent of the contact area $A$ ($\tau$ being the shear stress). $\mathbf{F}_{fric}$ is proportional to the applied normal force $\mathbf{F}_n= \overline{\sigma_{eff}} A$ through the constant $ \mu$ ($\overline{\sigma_{eff}}$ corresponds to the effective normal stress). We thus have:
\begin{equation}
\mu = \frac{\mathbf{F}_{fric}}{\mathbf{F}_n } = \frac{\tau}{\overline{\sigma_{eff}}}
\label{eqn:friction_force1}
\end{equation}
Let us now consider an object of mass $M$ placed on a table. The force $\mathbf{F} _ {n} = Mg $ is, therefore, normal to the surface. We apply a tangential force $ \mathbf{F}_{t} $ parallel to the surface of the table. If the object is initially at rest, a motion may be produced if a force $\mathbf{F}_{t}$, greater than  $\mathbf{F}_{fric}$, is applied. In this case, the coefficient $ \mu_{s} $ is called the coefficient of static friction. 
\begin{linenomath*}\begin{equation}
\mathbf{F}_{fric}=\mathbf{F}_{s} = \mu_s \mathbf{F}_n
\label{eqn:friction_force2}
\end{equation}\end{linenomath*}
Now, if the object is displaced at a finite velocity over the surface, it has been experimentally found that the frictional force is also proportional to the normal force, through the coefficient $ \mu_{d} $, called the coefficient of dynamic friction:
\begin{linenomath*}\begin{equation}
\mathbf{F}_{fric}=\mathbf{F}_{d} = \mu_d \mathbf{F}_n
\label{eqn:friction_force3}
\end{equation}\end{linenomath*}
Early experiments showed that the coefficient of static friction is different from the coefficient of dynamic friction  \cite{rabinowicz1958}.
Static friction has the property of increasing logarithmically with time, and dynamic friction depends on the velocity $ V$.

\begin{figure}
  \begin{center}
    \includegraphics[width=0.9\textwidth,angle=0]{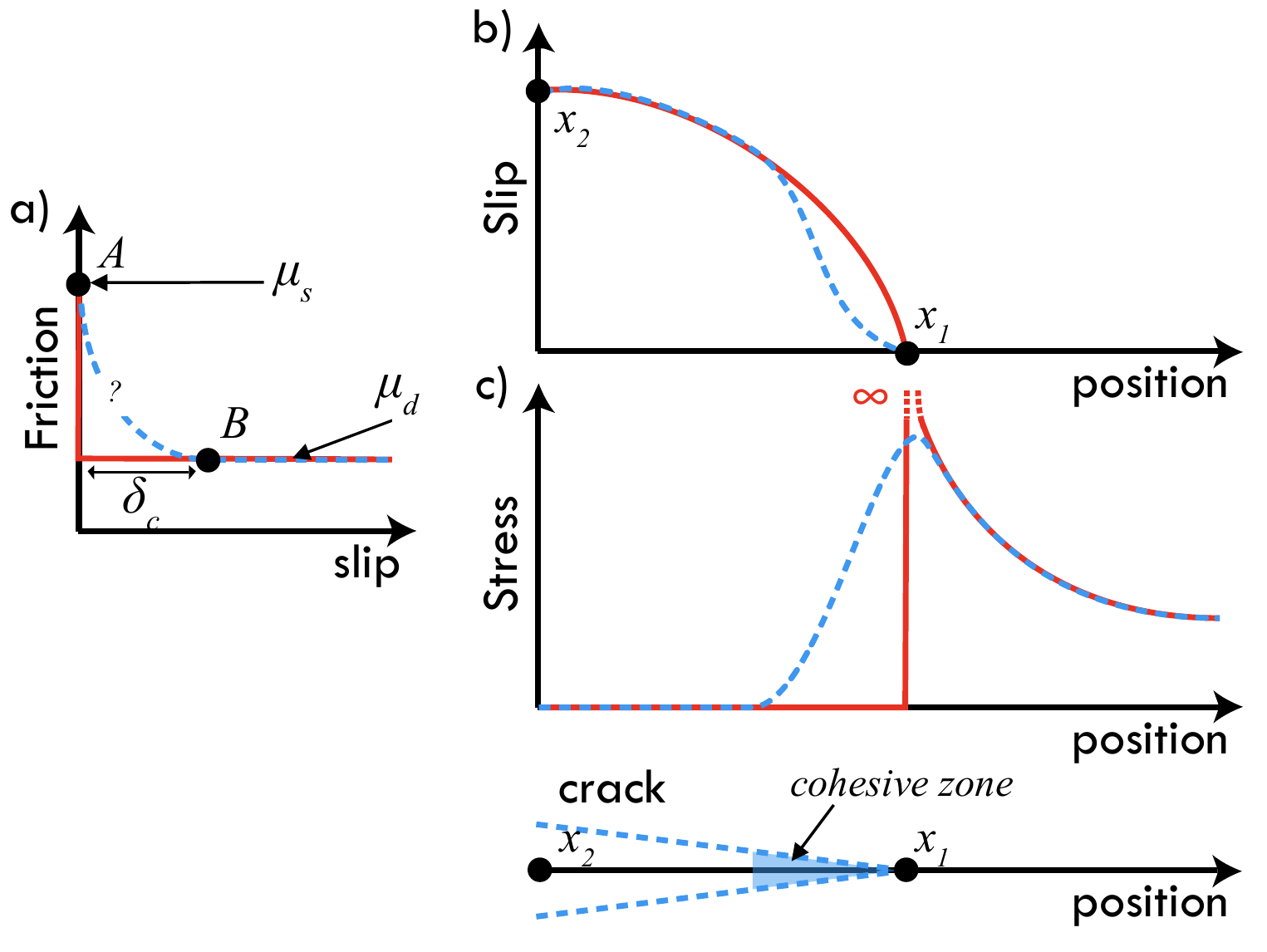} 
  \end{center}
   \caption{\small \textit{Comparison between the rupture model hypothesizing linear elasticity (red curve) and the cohesive zone model (dotted blue curve). a) Coefficient of friction in terms of the quantity of slip. b) Quantity of slip in terms of the position along the fracture. The point $x_1$ is in the position $A$ on the friction curve and the point $x_2$ is at position $B$. c) Stress field close to the rupture front. }\normalsize} 
   \label{fig:cohesive}
\end{figure}

From the classic work carried out by Kostrov \cite{kostrov1964, kostrov1966} and Eshelby \cite{eshelby1969}, it soon became clear that friction also played a fundamental role in the initiation, rupture development and `healing' of faults. The classic Amonton-Coulomb model, however, led to an impasse. Among other physical problems, it postulated the hypothesis of an instantaneous modification of the coefficient of friction, from its static value to its dynamic value. This brings in singularities (infinite stresses) at the rupture front (red model in figure~\ref{fig:cohesive}). 

This model lacks a scale of length that makes it possible to define a finite quantity of energy released at the rupture front. There are two possible options. One consists of defining the characteristic quantity of slip (between the two surfaces) required to move from static friction to dynamic friction. The other consists of introducing a characteristic time in which friction decreases from $\mu_s$ to $\mu_d$. In this second case, a scale of characteristic length emerges when the characteristic time is related to the slip velocity. 
For example: to explain his experiments on friction, Rabinowicz \cite{rabinowicz1958} introduced the concept of a "critical distance" $d_c$ during which the gap between the static friction and the dynamic friction is closed. He related this critical distance to the velocity, $V = D_c / t_{w}$. Here $ t_{w} $ is called \textit{weakening time}. 

In general, the laws called \textit{weakening friction laws} were thus developed to reproduce seismic behavior. We speak of \textit{weakening} because the friction reduces with the slip (or rate of slip) and these laws can thereby produce instabilities \cite{bocquet2013, zhuravlev2013, romanet2017b}. This ingredient is required to anticipate seismic velocities (m/s) in the models. We will now present the most used models in the following sections.

\subsection{Slip weakening friction law}
\label{subsec:SWL}

In fracture mechanics, the model where friction weakens with distance, also known as the \textit{cohesive zone model}, postulates that:

\begin{itemize}[
  align=left,
  leftmargin=2em,
  itemindent=0pt,
  labelsep=0pt,
  labelwidth=2em
]
\item the rupture process, which causes the shift from static friction to dynamic friction, is confined to the fracture plane,
\item inelastic deformation begins when the stresses on the rupture front reach a certain critical level,
\item we reach the value of the coefficient of dynamic friction when the displacement on the fracture plane exceeds a critical value $\delta_ {c}$ \cite{leonov1959, barenblatt1959, dugdale1960}.
\end{itemize}
This law was introduced in the context of a study of tension fractures, in order to solve the problem of singularities coming up (infinite stresses) on the rupture front (blue model in figure~\ref{fig:cohesive}).

The slip weakening friction law was introduced by Ida \cite{ida1972a} and Andrews \cite{andrews1976} to model dynamic ruptures for 2D models, and by Day \cite{day1982a} for 3D models. This is analogous to the cohesive zone model, but for mode II fractures, that is, for shear fractures. 
In this law, the slip is zero until the shear stress $\tau$ reaches a maximum value (elasticity limit) that will be denoted by $\tau^{s}_{f}$. Once this stress is attained, the slip starts and the resistance to the sliding $\tau_{f}$ decreases linearly until the value $\tau^{d}_{f}$, i.e., when the plane has slipped with a critical value $\delta_{c}$:
\begin{linenomath*}\begin{equation}
\label{eqn:SWL}
\tau_f(\delta) = \begin{cases}
(\tau^{s}_{f}-\tau^{d}_{f})\left(1-\dfrac{\delta}{\delta_{c}}\right) + \tau^{d}_{f} ~~~~~~~~&;\delta<\delta_{c}\\
\tau^{d}_{f} ~~~~~~~~&;\delta>\delta_{c}\\
\end{cases}
\end{equation}\end{linenomath*}
If this law is combined with the Amonton-Coulomb law (equation~\ref{eqn:friction_force1}), we have:
\begin{linenomath*}\begin{equation}
\label{eqn:SWL2}
\tau_f(\delta) = \begin{cases}
\left[(\mu_{s}-\mu_{d})\left(1-\dfrac{\delta}{\delta_{c}}\right) + \mu_{d}\right]\overline{\sigma_{eff}} ~~~~~~~~&;\delta<\delta_{c}\\
\mu_{d}\overline{\sigma_{eff}} ~~~~~~~~&;\delta>\delta_{c}\\
\end{cases}
\end{equation}\end{linenomath*}
where $\mu_d<\mu_s$. In their article, Palmer and Rice \cite{palmer1973} presented a law that is very close to this 
for which they could derive a complete analytical solution for the rupture front. They showed that this law made it possible to regularize the numerical model by distributing the stresses and the slip over a distance controlled by the length scale in the friction law.

\begin{figure}
\centering
\includegraphics[width=0.6\textwidth]{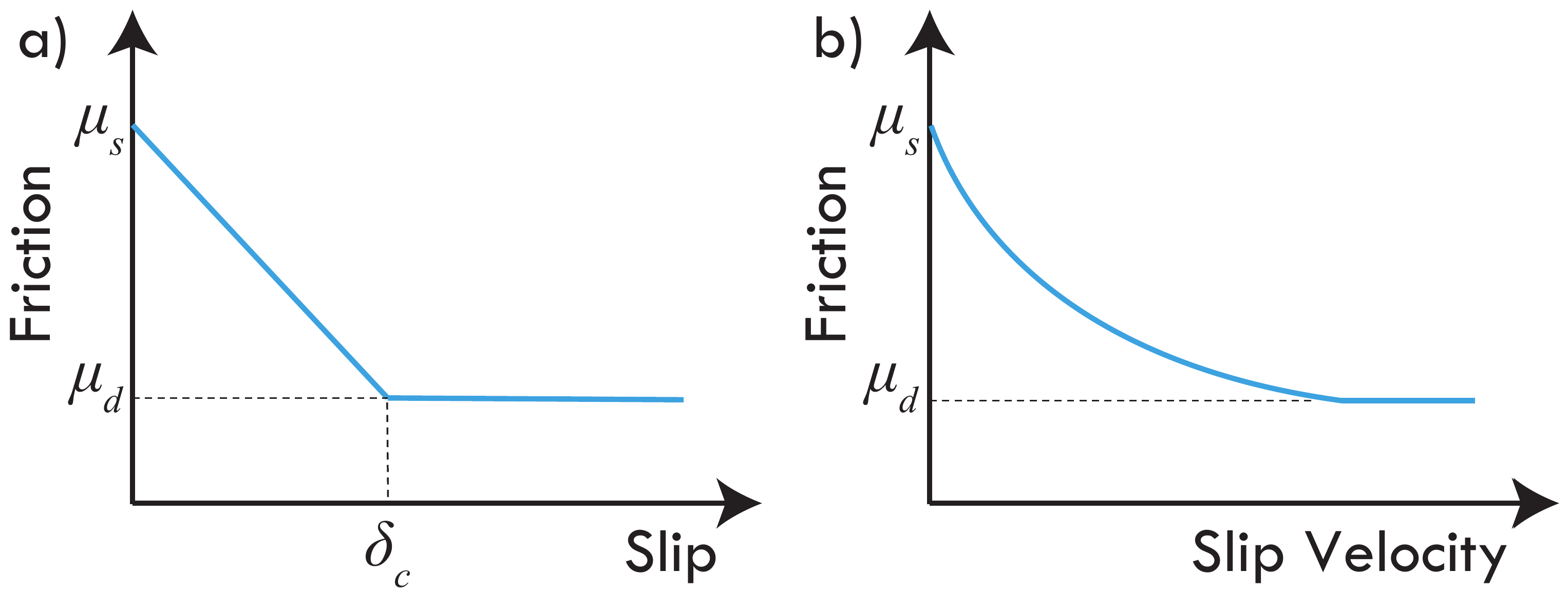}
\caption{Schematic illustration of (a) the \textit{slip weakening friction law}, (b) the \textit{velocity weakening friction law}}
\label{slipweak}
\end{figure}

\noindent A few nuanced but important points with respect to the slip weakening law:
\begin{enumerate}[leftmargin=*,labelindent=0pt,label=\roman*.]
\item This friction law describes the start and growth of a seismic rupture. The more the fault slips, the weaker its resistance. If the shear stress on the fault, $\tau$, is uniform, then this law implies that the fault will continue to slip indefinitely until $\tau< \tau_{f}$. This does not match the observations. There are therefore two possibilities: either $ \tau $ is heterogeneous along the the fault due to its geometric complexity (branches, non-linear plane, fault jump etc.) or related to past earthquakes. The second possibility, since faults have finite length, is that the rupture stoped because the earthquake ruptured the entire slip plane. Consequently, when it arrived at the geometric limit of the fault, the friction resistance $ \tau_{f} $, is infinite by definition. For most small earthquakes it seems likely that the first case is the applicable one. For larger earthquakes it may be assumed that the second case is applicable.
\item This law does not explain how the next earthquake will occur. 
Following an earthquake, the entire fault plane that reruptured should, logically, have a shear stress equal to the dynamic friction multiplied by the effective normal stress i.e., $\tau= \tau^{d}_{f} = \mu_{d}\overline{\sigma_{eff}}$. Further, for the nucleation and propagation of the next earthquake, $\tau$  must again increase and reach the value $\tau^{s}_{f}$. We talk about a fault plane `healing', but the  slip-weakening law does not allow this. It is thus well-suited to model a single rupture, but not to simulate the seismic cycle, where inter-seismic periods and earthquakes succeed one another over a long period of time.
\item If we go back to law~\ref{eqn:SWL}, but $\mu_s<\mu_d$, we will then have an increase in friction with the slip, which does not produce instabilities. We then talk of \textit{slip-hardening} behavior, which leads to `creep' type behavior.
\end{enumerate}

\subsection{Rate weakening friction law}
\label{subsec:RWL}

In order to respond to the problem of the fault plane 'healing',  i.e., to allow the shear value $\tau$ to return to the value $\tau^{s}_{f}$, Burridge and Knopoff \cite{burridge1967} propose a new model. They base it on a key observation made in the laboratory: once the plane has slipped from the critical value $\delta_c$, the friction becomes a function of the slip rate $V$ :
\begin{linenomath*}\begin{equation}
\label{eqn:RWL}
\tau_f(V) = (\tau^{s}_{f}-\tau^{d}_{f})\dfrac{V_{0}}{V_{0}+V} + \tau^{d}_{f}
\end{equation}\end{linenomath*}
where $V_{0} $ corresponds to the characteristic slip velocity. When the slip velocity is much smaller than $ V_{0} $, the fault's resistance to slip corresponds to the static friction ($\mu_{s}$) multiplied by the effective normal stress ($\overline{\sigma_{eff}}$), i.e., $\tau^{s}_{f}$.
Conversely, when the slip velocity is much greater than $V_{0}$,  the fault's resistance to slip corresponds to $\tau^{d}_{f}=\mu_{d}\overline{\sigma_{eff}}$.
Therefore, during an earthquake, the resistance decrease as the slip velocity is large (of the order of 1 m/s). On the other hand, it rises again quickly as the slip on the fault slows down, when it reaches loading velocities of the order of a mm/year to cm/year.
This, this law can not only model an earthquake individually, but also model the entire seismic cycle. Burridge et Knopoff \cite{burridge1967} applied this friction law over a series of connected block-spring systems used as a proxy for an elastic medium hosting a fault (cf. section~\ref{subsubsec:patinressort}).

\subsection{\textit{Rate-and-state} type friction law}
\label{subsec:RSL}

\begin{figure}
\begin{center}
\includegraphics[width=1.0\textwidth]{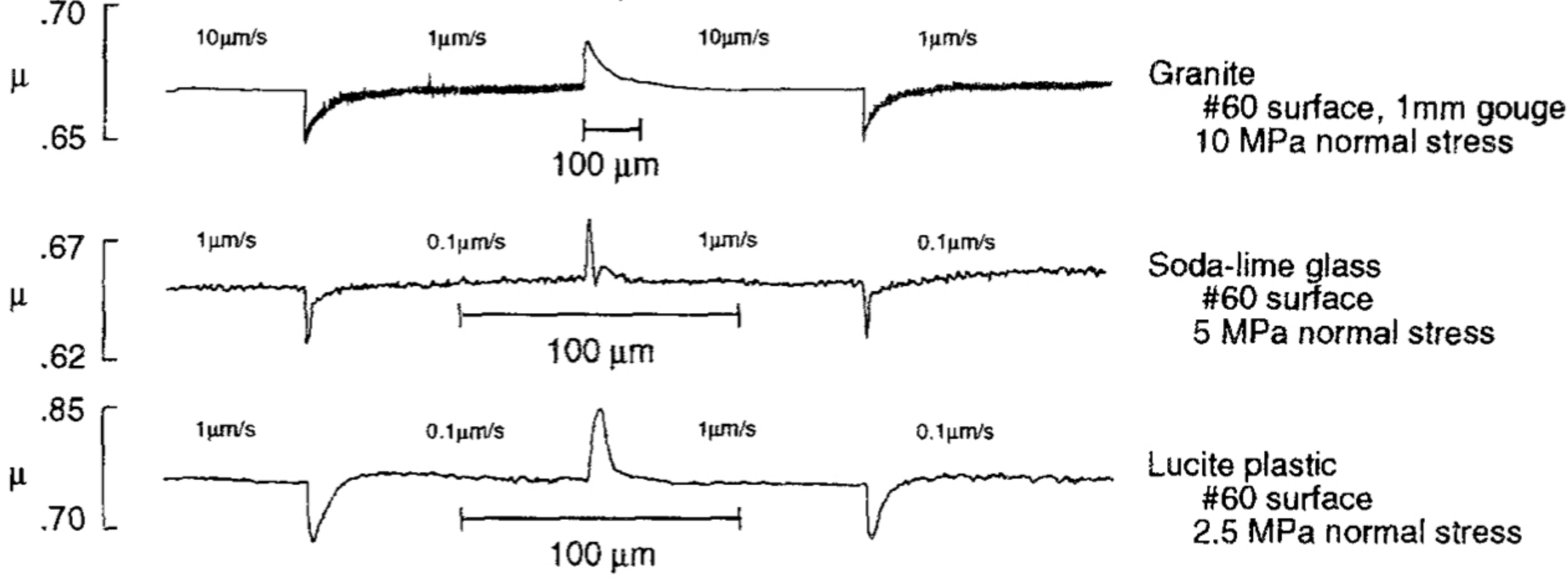}
\end{center}
\caption{Experiments on friction, by applying velocity jumps, for different types of materials, published by Dieterich in 1994 \cite{dieterich1994a}}
\label{fig:dieterich}
\end{figure}

\begin{figure}
\begin{center}
\includegraphics[width=1.0\textwidth]{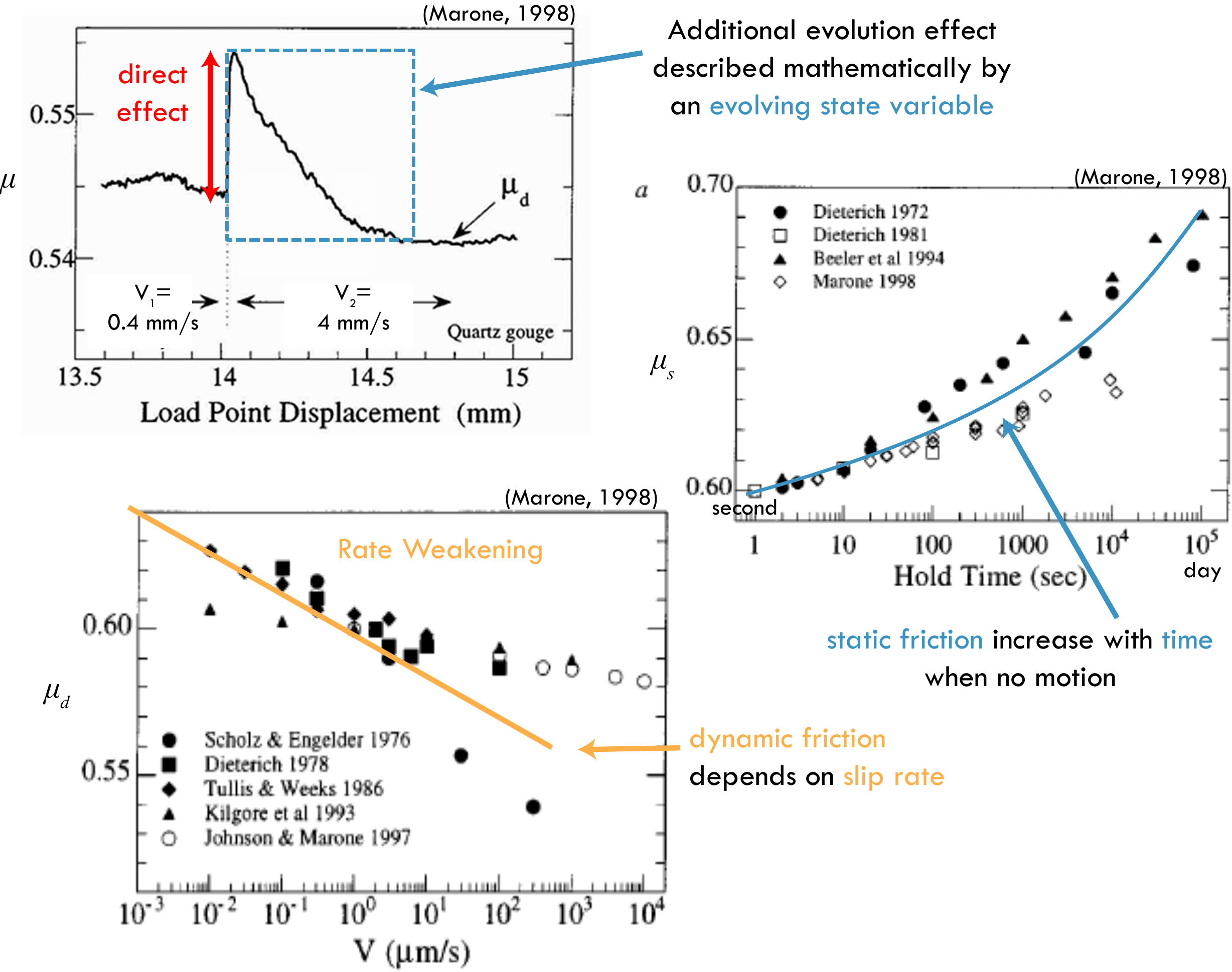}
\end{center}
\caption{Experiments on friction. Figures modified as per C. Marone  \cite{marone1998}}
\label{fig:marone}
\end{figure}

Continuing with the work started by Brace and Byerlee \cite{brace1966b}, new experimental protocols have emerged. In particular, researchers wished to explore the effect of the sudden change in velocity observed in nature, when there is a shift from aseismic velocities ($\sim$cm/yr) to seismic velocities ($\sim$m/s). Experiments with velocity jumps in the loading of the system were carried out (figure~\ref{fig:dieterich}). 
%
\smallskip
In his seminal 1998 paper, Chris Marone \cite{marone1998} offered an exhaustive review of these works. 
There are four key observations from this (figure~\ref{fig:marone}).
\begin{itemize}[
  align=left,
  leftmargin=2em,
  itemindent=0pt,
  labelsep=0pt,
  labelwidth=2em
]
\item A sudden change in slip rate first leads to a sudden increase in the coefficient of friction. This is called the direct effect.
\item A transient adjustment is then seen towards a new, stationary value of the coefficient of friction.
\item The coefficient of dynamic friction depends on the slip velocity.
\item The coefficient of static friction increases with time when there is no motion between the two surfaces in contact.
\end{itemize}

James H. Dieterich was the first person to propose an empirical law that could reproduce these observations both qualitatively and quantitatively \cite{dieterich1979a, dieterich1979b}. He based this, notably, on his own friction experiments, with velocity jumps, that involved two ground blocks of granodiorite. He also based it on his earlier experiments, demonstrating the coefficient of static friction increased with time \cite{rabinowicz1958}. He thus interpreted the decrease of the coefficient of friction with velocity as an effect of the reduction of the mean contact time. And so, in his friction law, the coefficient of friction goes from $\mu_s$ to $\mu_d$ over a distance  $D_c$, which relates the contact time $t$ to the slip velocity $V$ in the following manner: $V=D_c/t$. With this, he adopted an approach that was similar to that proposed by E. Rabinowicz (cf. section~\ref{subsec:muSmuD}). The law that he proposed made it possible to bring together the different coefficients of static and dynamic friction into a single coefficient, which depended on the slip rate. 
It was later refined by \cite{ruina1983}, through the introduction of a state variable $\theta$, which followed a law of evolution.
A common way to interpret $\theta$ is to relate it to the lifespan of the asperities present on the surfaces in contact.
The law was thus called the \textit{rate-and-state} law, due to the existence of this \textit{state} variable, and the dependence of the coefficient of friction on the velocity or \textit{rate}.

\begin{figure}
\begin{center}
\includegraphics[width=0.65\textwidth]{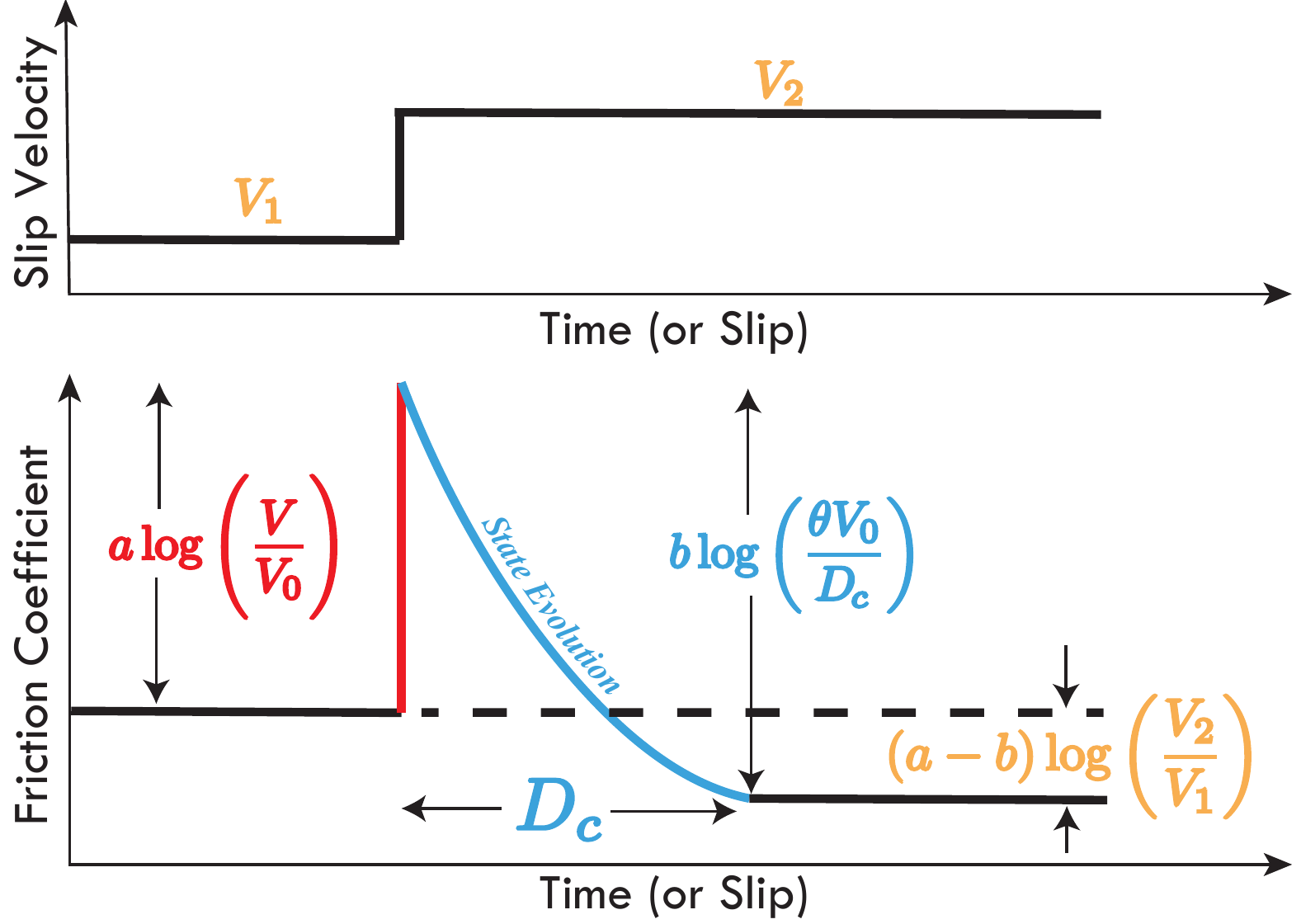}
\end{center}
\caption{Schematic illustration of the rate-and-state law}
\label{fig:ratestate}
\end{figure}

\noindent
A modern form of the rate-and-state law was given by \cite{marone1998}:
 \begin{linenomath*}\begin{equation}
 \label{eqn:RSL}
\tau_{f}(V,\theta) =  \left[\mu_0+a\log\left(\frac{V}{V_0} \right) +b\log\left(\frac{\theta V_0}{D_c} \right) \right]\overline{\sigma_{eff}}
\end{equation}\end{linenomath*}
By associating this either with a law called the \textit{aging law}:
\begin{linenomath*}\begin{equation}
 \label{eqn:RSL2}
  \dot{\theta} = 1 - \frac{\theta V}{D_c}
\end{equation}\end{linenomath*}
or with a state law called \textit{slip evolution}:
\begin{linenomath*}\begin{equation}
 \label{eqn:RSL3}
 \dot{\theta} = - \frac{V\theta}{D_c}\log \left(\frac{V\theta}{D_c} \right)
\end{equation}\end{linenomath*}
Here, $a> 0$ and $b$ are state parameters, of an order of magnitude of $\sim 10^{-2}$ , associated, respectively, with the direct effect and the transient change in the coefficient of friction (Figure~\ref{fig:ratestate}).  $f_{0}$ corresponds to the refernce coefficient of friction at the reference velocity $V_{0}$.

At constant slip velocity, $V$, the coefficient of friction ad the state variable evolve toward a stationary value, $f_{ss}$ and $\theta_{ss}$. It is thus possible to rewrite the rate-and-state law as follows:
 \begin{linenomath*}\begin{equation}
\label{fig:RSLss}
\theta_{ss} = D_{c}/V ~~~~~\&~~~~ f_{ss} = f_{0} + (a-b)\log{\frac{V}{V_{0}}}
\end{equation}\end{linenomath*}
Thus, when $(a-b)<0 $, the coefficient of friction decreases with the increase in slip velocity. We then speak of a \textit{rate-weakening} material. If $ (a-b)> 0 $ then a \textit{rate-strengthening} behavior is obtained.

Today, none of the state laws (equations~\ref{eqn:RSL2} and \ref{eqn:RSL3}) reproduce the full set of experimental data. The slip evolution law does not reproduce the logarithmic time dependence of the coefficient of static friction (figure~\ref{fig:marone}).  If $\dot{\delta} = 0 $, $\theta$ does not evolve over time. This is probably why the models tend to favor the \textit{aging law} \cite{ampuero2008a}. 
However, this law offers a non-symmetric response according to which a positive (increase) or negative (decrease) velocity jump is introduced \cite{blanpied1998, ampuero2008a}.
Several modifications were proposed to improve the state law. For example, by introducing a dependency for the normal stress \cite{linker1992}, by proposing a completely different evolution of the parameter $\theta$ \cite{perrin1995, kato2001}, or by adding a dependency to the shear rate \cite{bhattacharya2015} . However, none of these laws led to a consensus. 
On the other hand, other promising modifications made it possible to come close to observations made in nature (cf section~\ref{subsec:addweakening}). Some of those include additional friction mechanisms that increase friction through dilatancy \cite{segall1995, segall2012b}, or lead to a decrease in effective friction through the pressurization of pore fluids \cite{rice2006, schmitt2011}.

\section{Modeling fault behavior: the `spring-block slider' model }
\label{sec:modelfaille}

In the brittle part of the crust, the deformation is essentially accommodated along faults in response to the tectonic plate movement in the earth's crust. Along these faults two main behaviors are observed: either the fault creeps continuously at a velocity comparable to the plate velocity (mm/yr to cm/yr), or it remains locked for years, or even centuries, and slips suddenly in a very short time, of the order of several seconds, thus resulting in an earthquake. An earthquake of magnitude $M_w$ 4-5 corresponds to an average slip of a few centimeters, a $M_w$ 7 corresponds to a slip of a few meters, and a $M_w$ 9 to 10 to a slip of 20 meters. It is thus observed that slips of the order of m/s, cause destructive seismic waves that propagate in the surrounding medium. A simple analogy to represent the behavior of faults on the Earth's surface is the `spring-block slider' model (Figure~\ref{fig:springslider}), which is described in the following section.

\subsection{Modeling the slip on a fault: creep or earthquake}
\label{subsec:modelslip}

\subsubsection{Block-spring model}
\label{subsubsec:patinressort}

\begin{figure}
  \begin{center}
    \includegraphics[width=0.9\textwidth,angle=0]{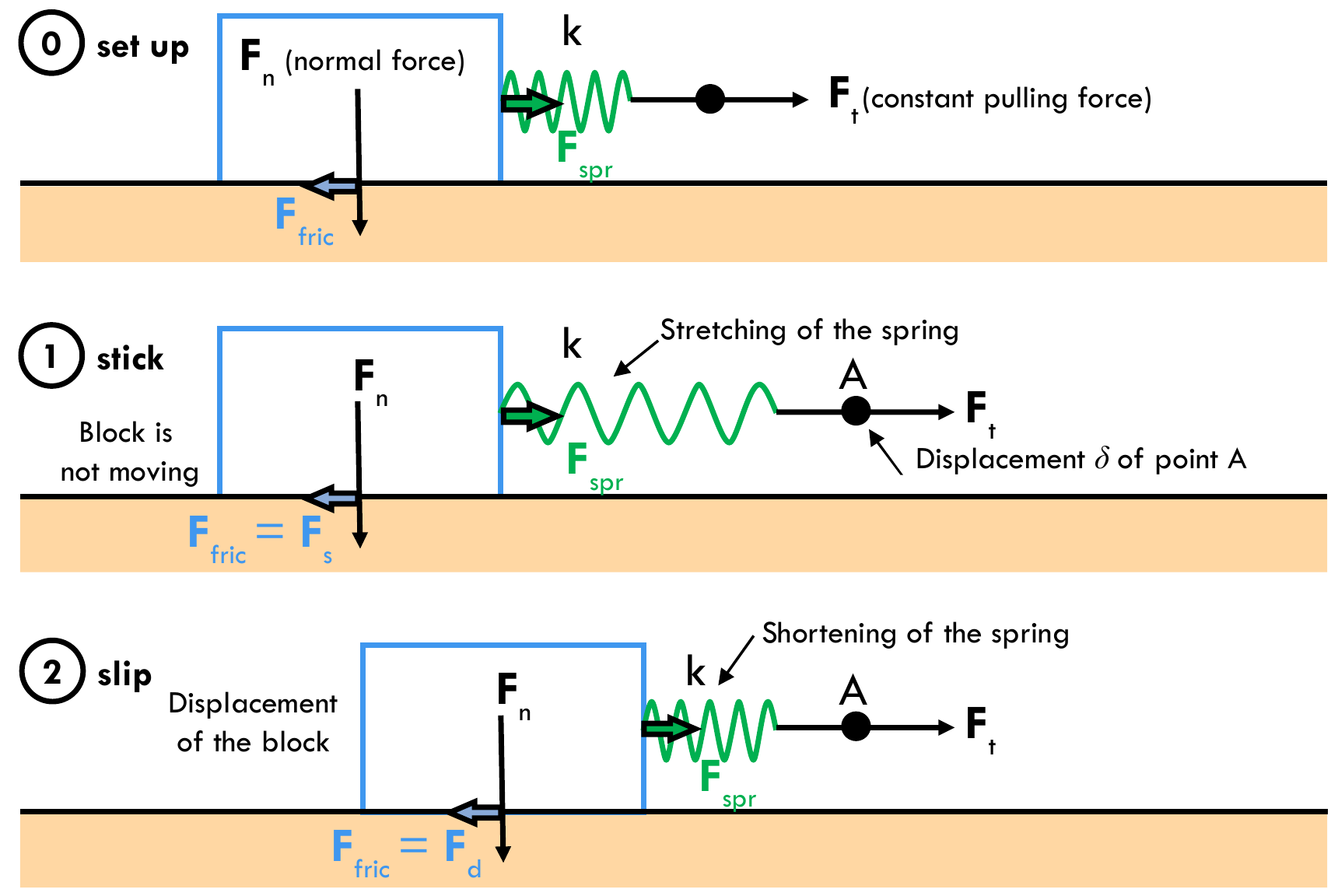} 
  \end{center}
   \caption{\small \textit{Spring Block slider model}\normalsize} 
   \label{fig:springslider}
\end{figure}

In the spring-block slider model, the force that pulls on the spring attached to the block in a constant manner represents the plate motion. The stiffness constant $k$ of the spring represents the rock's elastic properties, the weight of the spring, the compression and basal friction of the block, the friction of the fault plane (Figure~\ref{fig:springslider}).
There is therefore competition between the shear force pulling the block, $\mathbf{F}_{spr}$, and the force of the friction that resists the shear force, $ \mathbf{F}_{fric}$, defined as follows:
\begin{linenomath*}\begin{equation}
\label{eqn:Fspring}
\mathbf{F}_{spr} = \tau \times A = k \times x
\end{equation}\end{linenomath*}
\begin{linenomath*}\begin{equation}
\label{eqn:Ffriction}
\mathbf{F}_{fric} =\mu  \times \overline{\sigma_{eff}} \times A  = \mu \times \mathbf{F}_{n}
\end{equation}\end{linenomath*}
To recall: $\tau$ is to the shear stress, $A$ is the contact area,  $k$  is the spring's stiffness coefficient, $\overline{\sigma_{eff}}$ is the effective normal stress, and $\mu$ is the coefficient of friction. Depending on the law applicable to $\mu$, for example \textit{slip-hardening} or \textit{slip-weakening}, `creep' or `earthquakes' can be reproduced as observed in nature (cf. section~\ref{subsec:SWL}) . 

In the case of faults that produce earthquakes, we speak of \textit{stick-slip} behavior. That is, alternating between long periods where the fault does not move but stress accumulates (stick) and periods where the accumulated stress exceeds the fault's resistance to slip, which results in a slip displacement.

\subsubsection{Earthquake and instability condition}
\label{subsubsec:instability}

By applying a slip-weakening law to the block-spring model, it is therefore possible to reproduce stick-slip behavior and deduce the instability condition that will lead to a rapid, `earthquake' type slip. 

\begin{figure}
  \begin{center}
    \includegraphics[width=0.9\textwidth,angle=0]{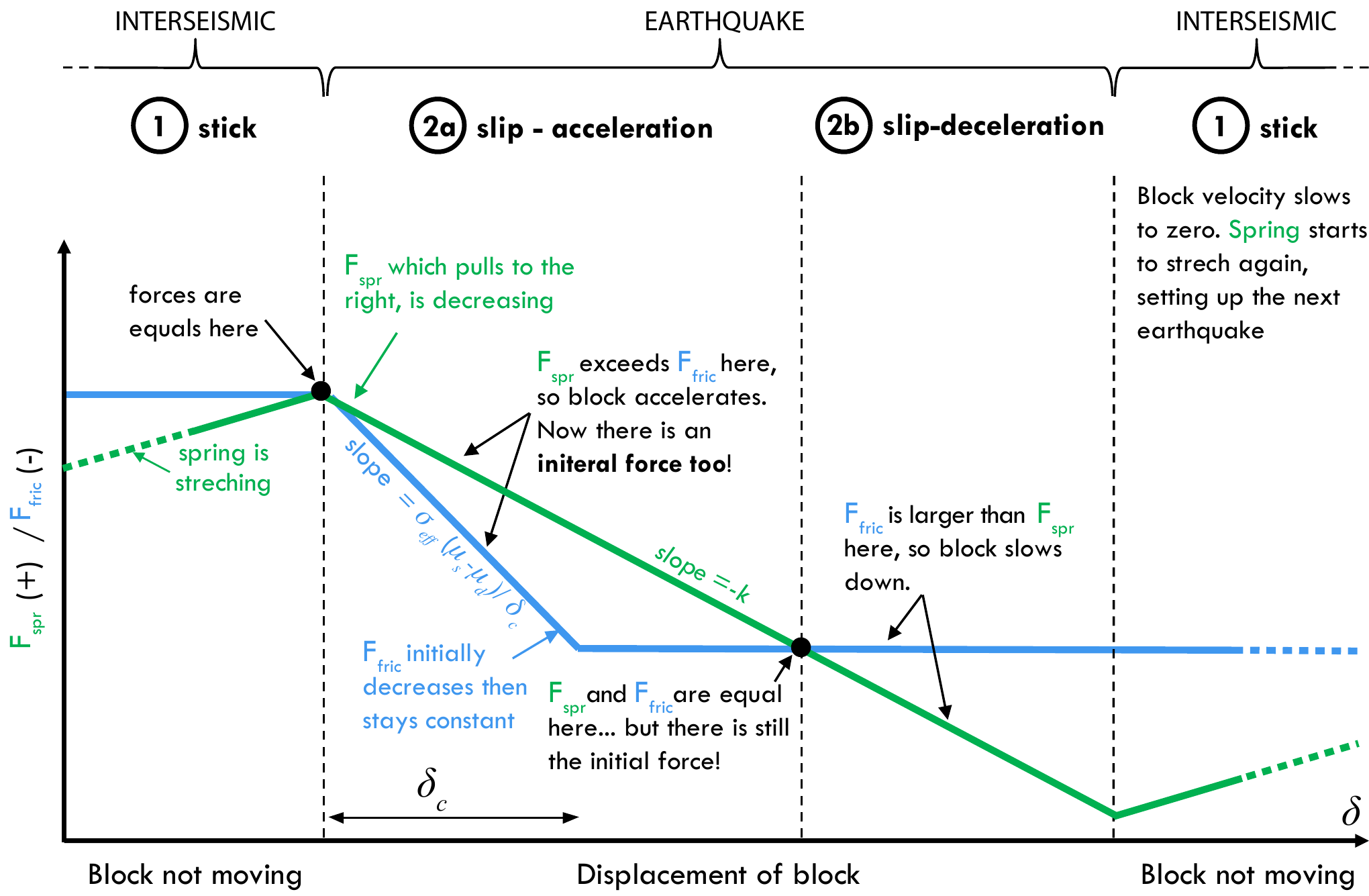} 
  \end{center}
   \caption{\small \textit{Balance equation of forces for the block-spring model with a slip-weakening friction law}}\normalsize 
   \label{fig:forcebalanceSWL}
\end{figure}

Initially, the spring is pulled over a distance $x$ but the block does not move (phase 1 in figures~\ref{fig:springslider} and~\ref{fig:forcebalanceSWL}). We thus have:
\begin{linenomath*}\begin{equation}
\label{eqn:stick}
F_{spr}+F_{fric}=0
\end{equation}\end{linenomath*}
Next, when the shear stress, $\tau$, which is equal to the fault's resistance to slip, $\tau^{s}_f=\mu_s\overline{\sigma_{eff}}$, the block begins to move. Since the block slips in the direction parallel to  $\mathbf{F}_{spr}$, this force decreases, just as $\mathbf{F}_{fric}$ because we applied a slip-weakening type friction to the model (cf eq.~\ref{eqn:SWL}). When $\mathbf{F}_{spr}$ exceeds $\mathbf{F}_{fric}$, the block accelerates (phase 2a. in figure~\ref{fig:forcebalanceSWL}). We therefore add an inertial force to equation~\ref{eqn:stick}.
\begin{linenomath*}\begin{equation}
\label{eqn:slip}
\mathbf{F}_{spr}+\mathbf{F}_{fric}=m\ddot{x}
\end{equation}\end{linenomath*}
When the coefficient of friction $\mu$ reaches its dynamic value $\mu_d$,  $\mathbf{F}_{fric}$ remains constant, while $\mathbf{F}_{spr}$ continues to decrease (phase 2b in figure~\ref{fig:forcebalanceSWL}). The block finally decelerates. After it completely stops, phase 1 (the stretching of the spring) resumes.

There is therefore an 1instability', i.e. an acceleration in slip, when $\mathbf{F}_{fric}$ decreases faster than $\mathbf{F}_{spr}$ during the slip. The instability condition is, thus, defined through the following relation, where $k$,  the stiffness of the spring, must be smaller than a critical value $k_c$:
\begin{linenomath*}\begin{equation}
\label{eqn:instability}
k<k_c=\left|\frac{\overline{\sigma_{eff}} (\mu_s-\mu_d)}{\delta_c}\right|
\end{equation}\end{linenomath*}
Conversely, creep is produced if $k>k_c$, i.e., if the system is 'rigid' (a high $k$) or if the normal stress is low.

\subsubsection{Representation of a subduction zone.}
\label{subsubsec:modelsubduction}
\begin{figure}
  \begin{center}
    \includegraphics[width=0.9\textwidth,angle=0]{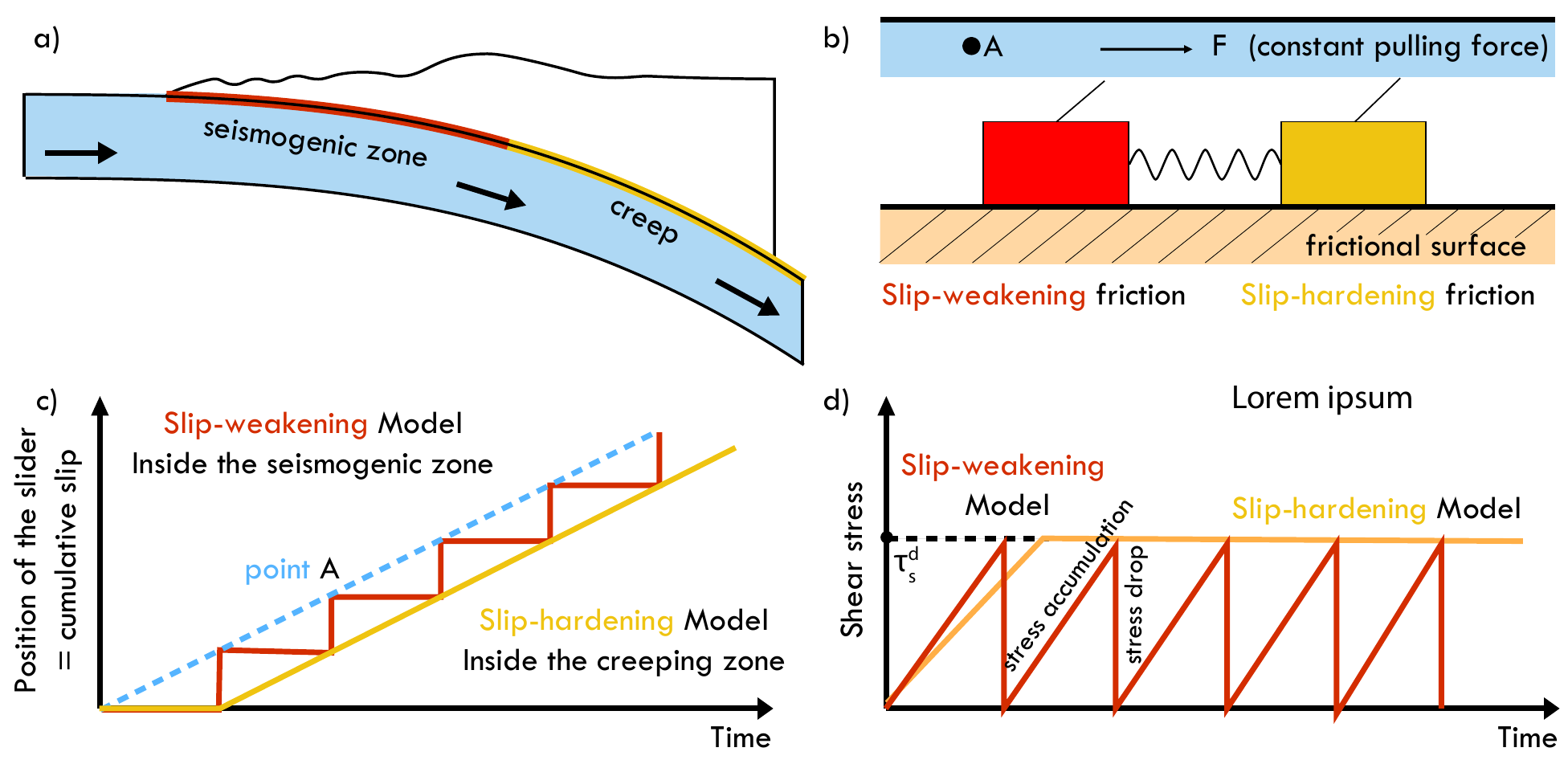} 
  \end{center}
   \caption{\small \textit{Modeling of a subduction using the block-spring method. a) chematic representation of a subduction. b) Conceptual model. c) Accumulation of slip over time. c) State of shear stress over time.}\normalsize} 
   \label{fig:subductionSWL}
\end{figure}

A simple way of representing a subduction zone, therefore, consists of combining several blocks, connected ot each other through springs, as proposed by Burridge and Knopoff in 1967 \cite{burridge1967}. The aseismic zone at depth is represented by a block whose basal friction responds to a slip-hardening law, and the seismogenic zone is represented by a block whose basal friction follows a slip-weakening law (figure~\ref{fig:subductionSWL}a and b). Researchers then observed that for the seismogenic zone, the slip accumulates in `steps' (figure~\ref{fig:subductionSWL}c). This is expressed by jagged variations in the shear stress, which is accumulated over long periods of time and then released in a few seconds (figure~\ref{fig:subductionSWL}d). We then speak of a \textit{stress-drop}. 
For the aseismic zone, after going through a plateau, which corresponds to the time required for the shear stress to reach the block's value of resistance to slip, i.e., $\tau^{s}_d$ (figure~\ref{fig:subductionSWL}d), the slip accumulates continuously and therefore there is indeed creep (figure~\ref{fig:subductionSWL}c). 

\subsection{Modeling the seismic cycle}
  align=left,
  leftmargin=2em,
  itemindent=0pt,
  labelsep=0pt,
  labelwidth=2em
]

\subsubsection{Shifting to the rate-and-state law}
\label{subsubsec:change4RSL}
\begin{figure}
  \begin{center}
    \includegraphics[width=0.9\textwidth,angle=0]{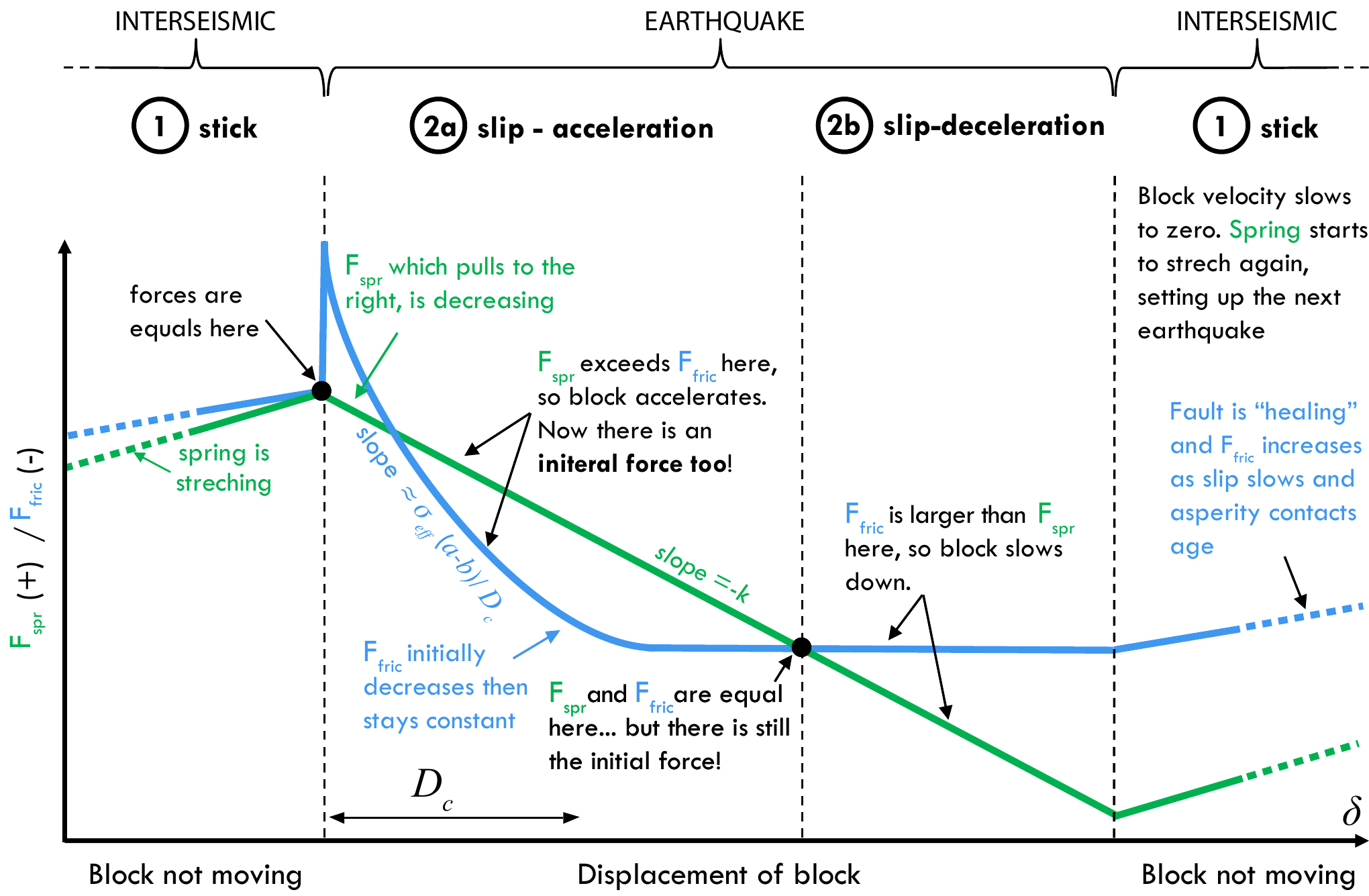} 
  \end{center}
   \caption{\small \textit{Assessment of forces for the block-spring model with a \textit{rate-weakening}  friction law (Rate-and-state law)}\normalsize} 
   \label{fig:forcebalanceRSL}
\end{figure}

As discussed in section~\ref{subsec:SWL}, while the earlier model makes it possible to reproduce the esential steps that lead to the seismic slip, it does not allow multiple events to be chained, since $\mu$ does not return to its static value $\mu_d$  (figure~\ref{fig:forcebalanceSWL}). 
On the other hand, the R\&S law, with the state variable $\theta$, takes into account the healing of the fault plane (figure~\ref{fig:forcebalanceRSL}).

If we go back to the spring-block slider model and replace the slip weakening friction law with a rate-and-state friction law, it is possible to derive a new instability condition. In this second case, during the acceleration phase (2a in figure~\ref{fig:forcebalanceRSL}), the slope of $F_{fric}$ is approximately equal to $\overline{\sigma_{eff}}(b-a)/D_c$. Consequently, for an instability, and potentially an earthquake, to be generated, we must have the following relation:
\begin{linenomath*}\begin{equation}
\label{eqn:instability2}
k<k_c\approx\left|\frac{\overline{\sigma_{eff}} (b-a)}{D_c}\right|
\end{equation}\end{linenomath*}

\subsubsection{Implications for the nucleation size of earthquakes}
\label{subsubsec:nucleation}
\begin{figure}
  \begin{center}
    \includegraphics[width=0.5\textwidth,angle=0]{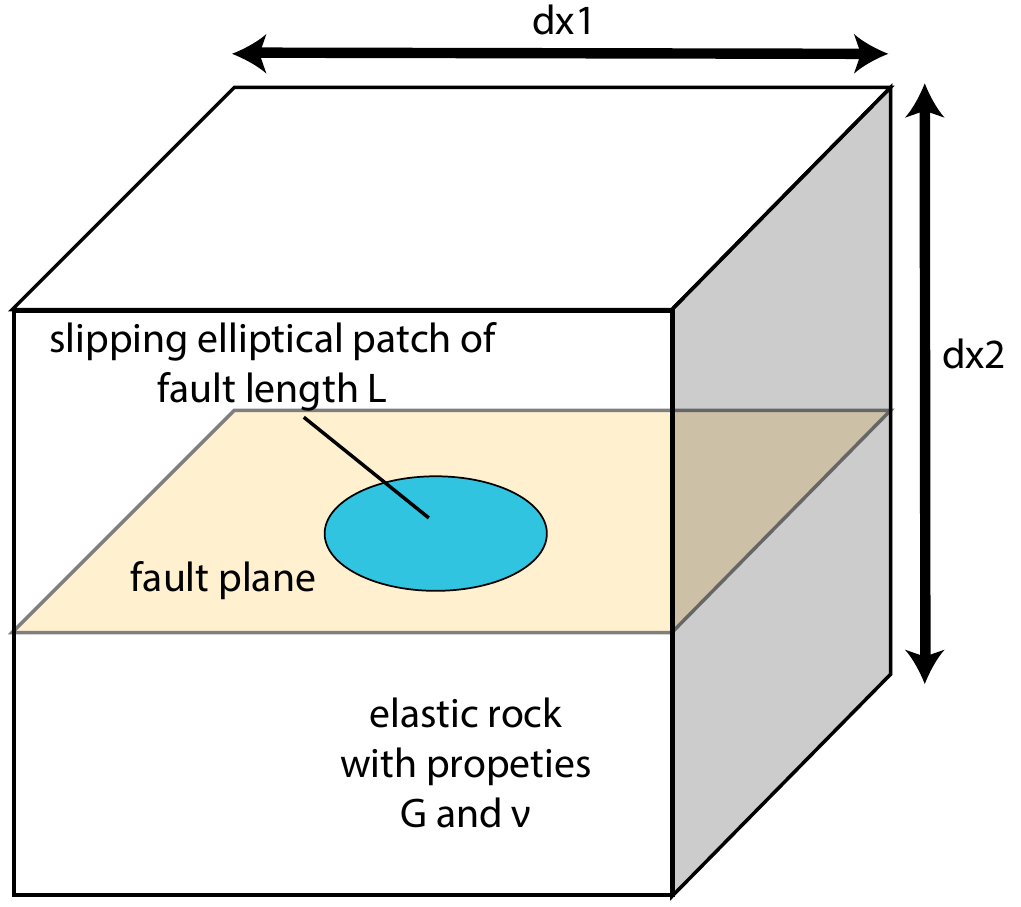} 
  \end{center}
   \caption{\small \textit{Nucleation model}\normalsize} 
   \label{fig:nucleation}
\end{figure}

To move from the spring-block slider model to a slightly more realistic Earth model with elastic behavior, we use elasticity to determine the $k$ value of an elliptical crack:
\begin{linenomath*}\begin{equation}
\label{eqn:kcrack}
k=\frac{G}{(1-\nu)L}
\end{equation}\end{linenomath*}
where $G$ is the shear modulus, $\nu$ is the Poisson's ratio and $L$ is the length of the zone that slips over the fault plane (figure~\ref{fig:nucleation}).
In this case, the instability occurs when the decrease in the frictional force is greater than the decrease in elastic force, and equation~\ref{eqn:instability2} is rewritten as: 
\begin{linenomath*}\begin{equation}
\label{eqn:instability3}
\frac{G}{(1-\nu)L}<k_c\approx\left|\frac{\overline{\sigma_{eff}} (b-a)}{D_c}\right|
\end{equation}\end{linenomath*}
Consequently, the zone that slips must be greater than a critical size $L_c$ in order to become unstable and generate earthquake nucleation:
\begin{linenomath*}\begin{equation}
\label{eqn:nucleation}
L>L_c\approx\left|\frac{D_cG}{(1-\nu)\overline{\sigma_{eff}} (b-a)}\right|
\end{equation}\end{linenomath*}

\subsubsection{Continuum model}
\label{subsubsec:modelcontinu}
In his seminal 1993 article \cite{rice1993}, J. R. Rice highlights the importance of moving from "spring-block slider" models to continuous medium models. He demonstrated, notably, that ``While the equations of Newtonian dynamics are solved exactly in these Burridge-Knopoff models, it has not been generally acknowledged that the dynamical solution for rupture along a chain of lumped masses,or a string of concentrated mass in the continuous limit, bears a presently uncertain relation to dynamical solutions for rupture along a fault embedded in a surrounding elastic continuum. For example, the response of B-K models to an instantaneous change in stress $\tau$ along the rupture is an instantaneous change in the acceleration $\partial^{2}\delta/\partial t^{2}$, but there is no instantaneous change in $\partial\delta/\partial t$.'' This is true, on the other hand, in continuum models. The other major drawback is ``Also, since there is no analogue to energy radiation as seismic waves in the normal implementation of the B-K models (an exception is the recent work of Knopoff et al. [1992]), all potential energy lost to the system during a rupture is fully account- able as frictional work; the same is not true for rupture in a continuum.''

It is therefore essential to highlight, in this text, that while the block-spring model makes it possible to qualitatively reproduce the phenomena observed in nature, it is essential to shift to a continuum model if we wish to develop robust numerical models. Interested readers can consult \textit{The mechanics of faulting: from laboratory to real earthquakes} \cite{Bizzarri2012}.

\section{A more complex physical reality }
\label{sec:stateofart}
  align=left,
  leftmargin=2em,
  itemindent=0pt,
  labelsep=0pt,
  labelwidth=2em
]

\subsection{Spatial and temporal variability in the slip mode on faults}
\label{subsec:var_glissement}

Until recently the deformation in fault zones, in the brittle part of the crust, was attributed either to earthquakes or to the slow, continuous slip during the inter-seismic period (creep) or post-seismic period. This latter phenomenon is called the \textit{afterslip} and corresponds to a logarithmic acceleration in the aseismic slip on the fault, which can be observed after large earthquakes. However, this paradigm of two 'extreme' behaviors is being questioned today. 

\begin{figure}
   \begin{center}
    \includegraphics[width=1.3\textwidth,angle=90]{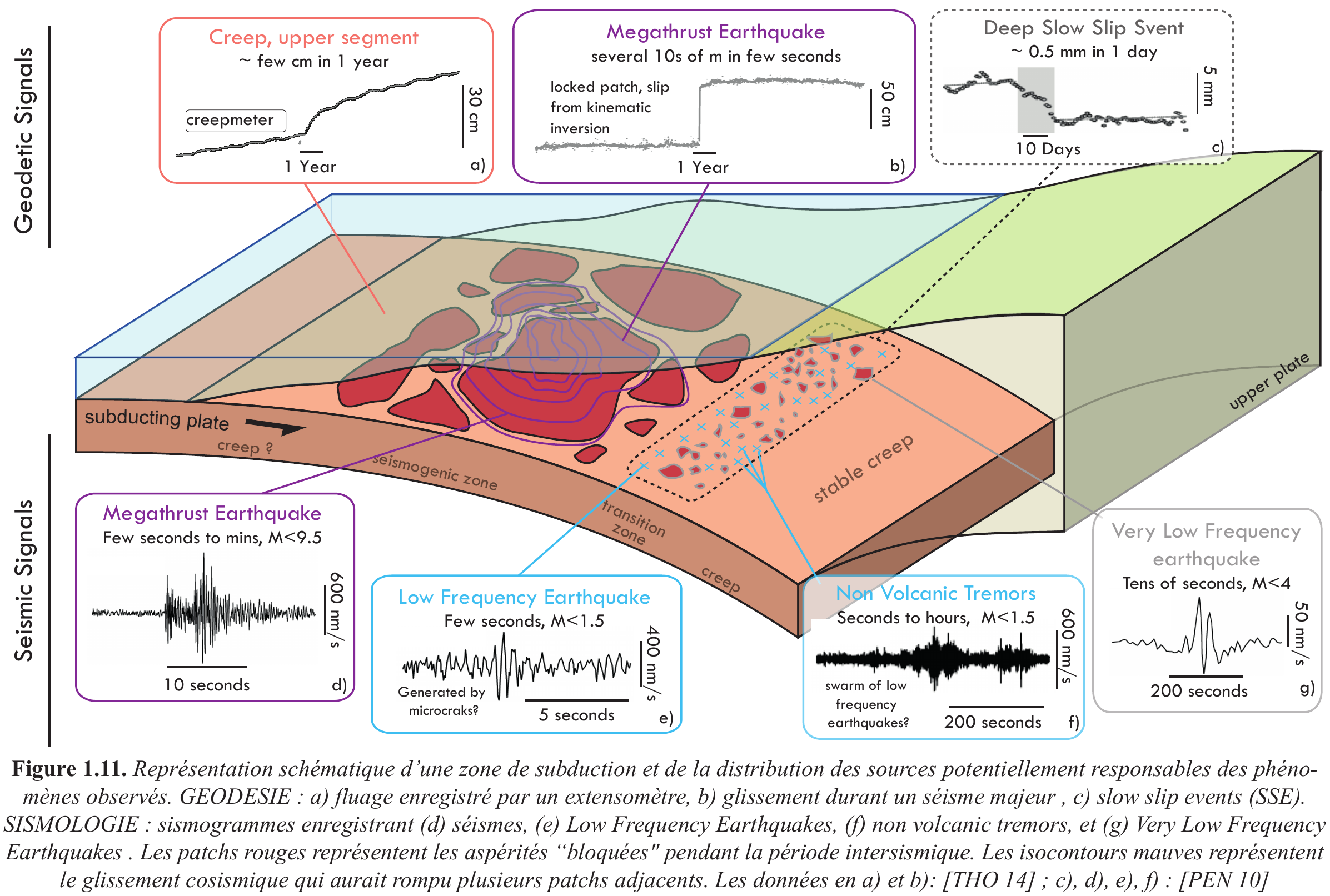} 
    \end{center}
\captionsetup{labelformat=empty} 
\caption{}
   \label{fig:fault_sismo}
 \end{figure}

Advances in technology and methodology in the field of geodesy and in seismology have significantly improved our capacity to measure deformation rates and given us higher resolutions. These observations have enabled us to document a large variability in the slip dynamic in the seismogenic zone (figure~\ref{fig:fault_sismo}). 
Faults may have chiefly seismic behavior, have a slow, stable slip \cite{Thomas2014b} or a transient slip \cite{Rousset2016}. In addition to this, one of the most significant discoveries in the last decade 
has been revealing the existence of 'slow earthquakes' (cf. Chapter 7). These encompass several phenomena.
 \textit{Slow slip events} rupture the fault very slowly over several hours or even days, at velocities that are higher than the inter-seismic creep (cm/yr), but slower than earthquakes, such that no detectable seismic waves are radiated \cite{Dragert2001}.  They are generally (though not always) accompanied by weak seismic signals of a long duration (a few minutes to a few weeks) called \textit{non volcanic tremors} \cite{Obara2002}. 
\textit{Low frequency earthquakes}, with a duration close to a second, and \textit{Very low frequency earthquakes}, which can last a hundred seconds, are commonly observed within \textit{non volcanic tremors} \cite{Ide2007,Ito2007}. 
As a result, it is known today that slip velocities on faults cover a continuum going from a millimeter per year to a meter per second  \cite{Peng2010}. This is therefore an essential parameter to take into consideration when modeling active faults. However, the physics behind the processes that govern this behavior is still unknown and is the subject of much active debate in the community.

In addition to the large range of deformation velocities there is a spatial and temporal variability in the slip mode. Contrary to what the schematic representation of figure~\ref{fig:fault_sismo} might suggest, the phenomena described here are not restricted to a specific depth. On some faults creep may be recorded over the entire seismogenic zone i.e. from the surface up to the maximum depth where earthquakes are observed \cite{Titus2006,Thomas2014b}. Further, while slow earthquakes were first located beyond the seismogenic zone \cite{Obara2002,Ide2007}, non volcanic tremors and slow slip events have recently been observed at epths of less than 10 km, as well as in the sub-surface \cite{Ito2006,Outerbridge2010}. 
Moreoever, geodetic data has shown that the seismic or aseismic behavior is not necessarily stable over time, and that the same zone may creep and slide seismically \cite{Johnson2012,Thomas2017a}. 
These observations lead to two hypotheses. (1) These different phenomena can occur under varied pressure/temperature conditions and/or result from various deformation mechanisms. (2) They correspond to particular mechanical and rheological properties, but which vary over time.  Consequently, they also vary over space, depending on what seismic cycle phase the observed site is undergoing.

\subsection{Additional mechanisms that can come into play during earthquakes}
\label{subsec:addweakening}

\begin{figure}
  \begin{center}
    \includegraphics[width=0.9\textwidth,angle=0]{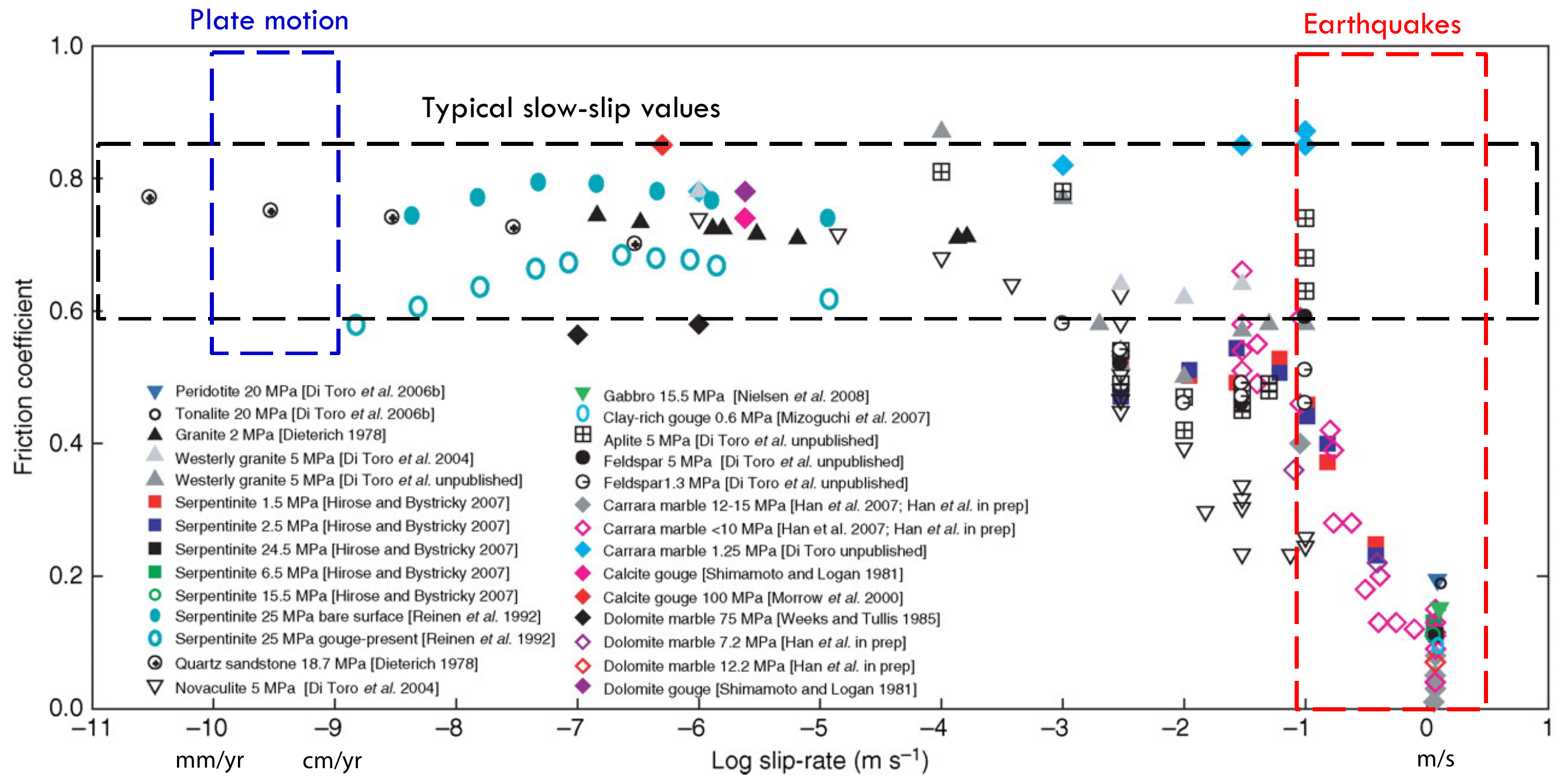} 
  \end{center}
   \caption{\small \textit{Dependence of the coefficient of dynamic friction, in a continuous regime, on the slip velocity. Figure modified as per \cite{Wibberley2008}. }\normalsize}
\label{fig:wibberley}
\end{figure}

The standard formulation of the rate-and-state law, (section~\ref{subsec:RSL}), allows a numerical reproduction of a large number of the phenomena discussed above. However, this formulation was based on slip velocity experiments ranging from  $10^{-9}$ to $10^{-3}$ m/s. While comparable to aseismic velocities ($10^{-10}$ to $10^{-9}$ m/s), they are still slow when compared to seismic velocities ($\sim1$ m/s). 
There is increasing experimental and theoretical proof that larger slip velocities and quantity of slip also come into play \cite{Lapusta2012}. This has the effect of drastically reducing the dynamic friction. Wibberley and co-authors \cite{Wibberley2008} have compiled laboratory values for different kinds of rocks and at different loading velocities (figure~\ref{fig:wibberley}).

The lack of experimental data on the properties of friction that are applicable to earthquakes is due to the difficulty of carrying out experiments in conditions similar to earthquakes. A laboratory experiment that would reproduce the conditions that exist during seismic slip would simultaneously involve high slip rates (1-10 m/s), with large displacements (0.1-20 m), a resulting effective normal stress (50-200 MPa), high pore pressure (0.4 to 1 times the normal stress) and high temperature (ambient temperatures of 100 to $300^{\circ} $C, but potentially as high as 1500$^{\circ}$ C in the slip zone). Although considerable progress has been made over the last decade, there is as yet no device that is capable of simultaneously responding to all these requirement. It is therefore necessary to compromise on one or more factors. Tullis and Schubert highlighted this difficulty and proposed a complete review of the processes that could lead to substantial reductions in the friction coefficient with respect to its typical experimental value of 0.6 \cite{Tullis2015}. The proposed mechanisms include:
\begin{itemize}[
  align=left,
  leftmargin=2em,
  itemindent=0pt,
  labelsep=0pt,
  labelwidth=2em
]
\item dynamic reduction in the normal stress or loss of contact due to the vibrations perpendicular to the interface, 
\item dynamic reduction in the normal stress due to the contrast in elastic properties, or permeability, on either side of the fault,
\item acoustic fluidization,
\item elasto-hydrodynamic lubrication, 
\item thermal pressurization of pore fluids, 
\item pressurization of pore fluids induced by the degradation of minerals, 
\item local heating/melting of the point of contact between the asperities, 
\item lubrication of the fault through fusion, in response to frictional processes,
\item lubrication of the fault through the creation of a thixotropic silica gel, 
\item superplastic deformation of fine grains. 
\end{itemize}
These highlight the difficulty of proving which mechanism is responsible for the observed experimental behavior and to design experiments that can clearly prove or refute a mechanism proposed in theory.
Nonetheless, since it is likely that one or more of these processes is activated at high slip rates, the rate-and-state law described in section~\ref{subsec:RSL} does not adequately reproduce this strong fall in the coefficient of dynamic friction. Indeed, for seismic velocities ($\sim1$ m/s) is a typical value for $(a-b)$ equal to $-0.005$, we obtain a $\mu_d$ of $\sim0.54$. Further, based on laboratory experiments, the effectve $\mu_d$, i.e., $\tau/\overline{\sigma_{eff}}$, can reach very low values (0  to 0.2) during co-seismic slip. 
This observation has many implications for our understanding of the mechanism of earthquakes: on the amplitude of the stress drop, on the propensity of earthquakes to propagate in pulse form, on the amplitude of ground movements, and on the orientation of stresses in the crust. N. Lapusta and S. Barbot propose two ways of modifying the \textit{rate-and-state} law to take into account these additional weakening mechanisms \cite{Lapusta2012}. Interested readers may refer to their publication for more details.



\subsection{Going beyond the elastic Earth model}
\label{subsec:couplage_THMC}

\begin{figure}
  \begin{center}
    \includegraphics[height=0.8\textwidth,angle=90]{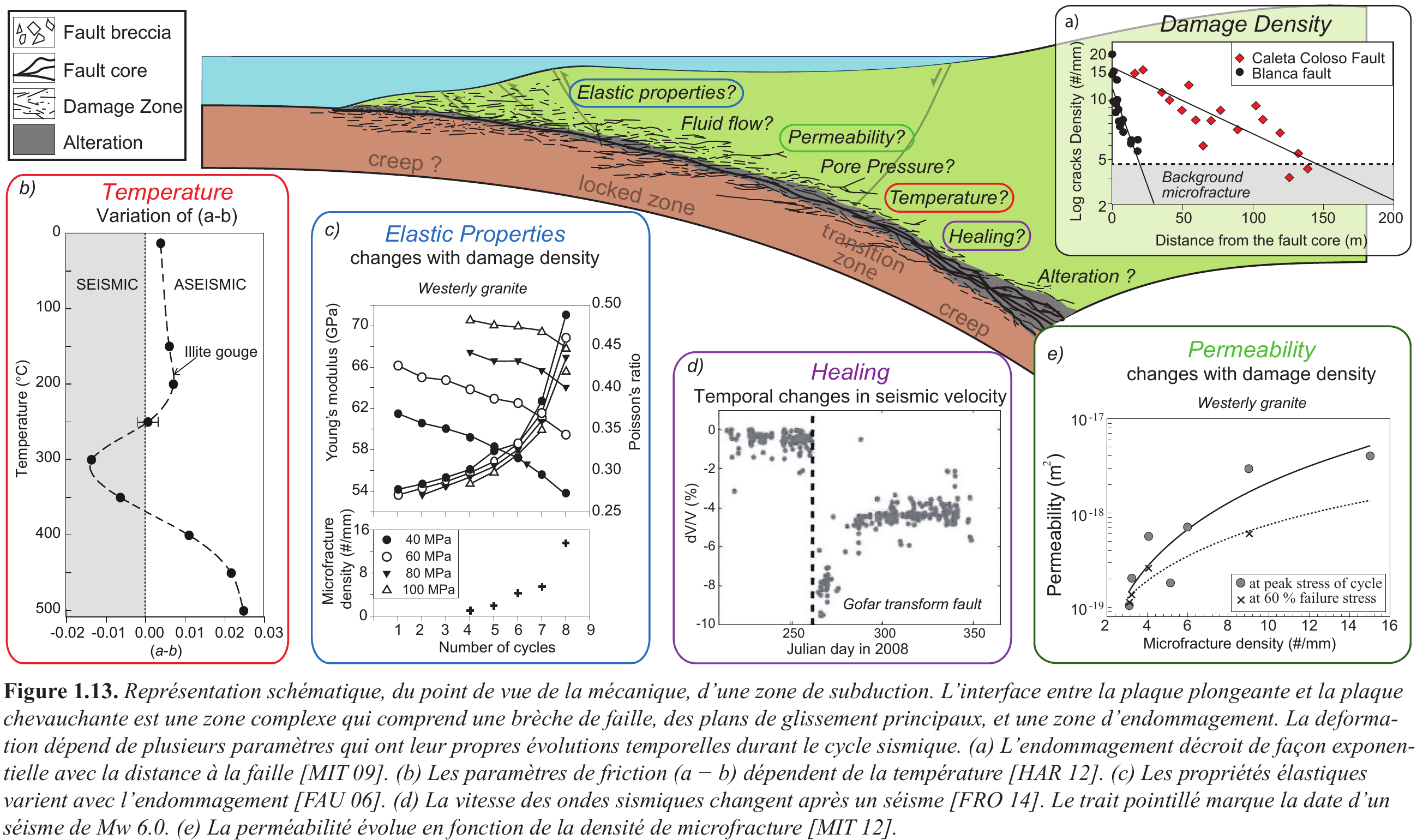} 
  \end{center}
\captionsetup{labelformat=empty} 
\caption{}
\label{fig:fault_complex}
\end{figure}

Many ground studied, geophysical observations, and laboratory experiments have highlighted the strong coupling that exists between the main rupture plane and the surrounding medium. .
The faults zones are not made up only of a major plane where the majority of slip occurs, but also make up a complex group, surrounded by a zone where surrounding rock is fractured intensively (figure~\ref{fig:fault_complex}). 
Seismic ruptures result in damage around the faults with an exponential decrease in the density of microfractures perpendicular to the main slip plane \cite{Anders1994,Mitchell2009}. The damage modifies the microstructure and changes the elastic properties of the rocks at the level of the fault breccia and in the adjacent medium \cite{Walsh1965, Walsh1965a, Faulkner2006}.
These changes, in return, modify the extension an dynamic of the rupture as well as the radiation of seismic waves \cite{Thomas2017b}. They also influence seismic processes during the post-seismic period, such as aftershocks, with the minimum size of the nucleation zone depending chiefly on the elastic modulus \cite{Rubin2005}. In their experimental study, \cite{Gratier2014} have also demonstrated that the co-seismic damage would promote aseismic slip through pressure-dissolution, thus explaining the afterslip recorded after large earthquakes.
%
%
\captionsetup[figure]{labelformat=empty}
\captionsetup[figure]{labelformat=default}
Co-seismic damage also increases permeability (figure~\ref{fig:fault_complex}e), which results in a variation in the fluid pressure \cite{Sibson1994} that modifies the fault's resistance to slip.
Geophysical observations suggest that this effect is transient (figure~\ref{fig:fault_complex}d), because a gradual and partial recovery of the elastic properties after the earthquake has been recorded \cite{Hiramatsu2005,Froment2014}.
This evolution is probably related to the healing of microfractures and faults through the precipitation of dissolved substances, 
products of alteration and/or the development of clayey minerals \cite{Mitchell2008}. In their model, \cite{Hartog2014} propose that the compaction through pressure-dissolution leads in turn to the recovery of seismogenic behavior.

Moreover, several studies have demonstrated the influence of the properties of the surrounding rock on the behavior of faults. Audet and co-authors have shown a direct relationship between the physical properties of the interlocking plate in the subduction zone and the recurrence of slow earthquakes  \cite{Audet2014}. In my microstructure study of Taiwan's longitudinal valley fault, Thomas and co-authors were able to demonstrate the aseismic behavior of the fault was controlled by inherited microstructure \cite{Thomas2014c}. \cite{Perrin2016} looked at the influence of the 'maturity' of the faults on the accumulation of slip. A study of 27 earthquakes concluded that the more damage the fault presents (mature fault), the greater the quantity of slip during an earthquake.


\section{Transition towards a new generation of models}
\label{sec:newmodel}

The usual way of looking at the fault restricts the deformation in the brittle part of the crust to slip along the interface (fault plane), loaded with creep at depth, whose behavior is controlled by its frictional properties \cite{Scholz1998}. According to these properties, when the resistance threshold is exceeded, the stress accumulated when the fault is locked is released through seismic slip or creep, or again during slow earthquakes. 
Further, as the previously cited studies have highlighted, while the behavior of the fault zones is intrinsically related to the properties of the main slip plane, it also depends on the properties of the surrounding rock. In parallel, the displacement on the faults induces a modification of the physical properties of the surrounding medium. these observations suggest the existence of a second `cycle' where the properties of the fault zone evolve with respect to the slip dynamic, which in turn influences the deformation mode.

However, the majority of models used today do not take this complex feedback into account. By attributing constant properties (pressure, temperature, petrology, microstructure) that do not evolve with deformation, we neglect to take into account how seismic/aseismic fault behavior is impacted by temporal variations of the physical properties of the volume and the interface.  
It is thus useful to develop a new generation of models that take into account spatial-temporal evolution of physical properties in fault zones. New models are being developed and have already shown the importance of these interactions from a seismic point of view \cite{Thomas2017b,Thomas2018a,Okubo2019}.


\begin{thebibliography}{67}
\expandafter\ifx\csname natexlab\endcsname\relax\def\natexlab#1{#1}\fi

\bibitem[Ampuero \& Rubin(2008)]{ampuero2008a}
Ampuero, J.-P. \& Rubin, A.~M., 2008.
\newblock Earthquake nucleation on rate and state faults--aging and slip laws,
  {\it J. Geophys. Res.\/}, {\bf 113}(B1).

\bibitem[Anders \& Wiltschko(1994)]{Anders1994}
Anders, M. \& Wiltschko, D.~V., 1994.
\newblock Microfracturing, paleostress and the growth of faults, {\it Journal
  of Structural Geology\/}, {\bf 16}(6), 795--815.

\bibitem[Andrews(1976)]{andrews1976}
Andrews, D.~J., 1976.
\newblock Rupture velocity of plane strain shear cracks, {\it J. Geophys.
  Res.\/}, {\bf 81}(B32), 5679--5689.

\bibitem[Audet \& Burgmann(2014)]{Audet2014}
Audet, P. \& Burgmann, R., 2014.
\newblock Possible control of subduction zone slow-earthquake periodicity by
  silica enrichment, {\it Nature\/}, {\bf 510}, 389--392.

\bibitem[Barenblatt(1959)]{barenblatt1959}
Barenblatt, G.~I., 1959.
\newblock The formation of equilibrium cracks during brittle fracture. general
  ideas and hypotheses. axially-symmetric cracks, {\it J. Appl. Math.
  Mech.-USSR.\/}, {\bf 23}(3), 622--636.

\bibitem[Bhattacharya et~al.(2015)Bhattacharya, Rubin, Bayart, Savage, \&
  Marone]{bhattacharya2015}
Bhattacharya, P., Rubin, A.~M., Bayart, E., Savage, H.~M., \& Marone, C., 2015.
\newblock Critical evaluation of state evolution laws in rate and state
  friction: Fitting large velocity steps in simulated fault gouge with time-,
  slip-, and stress-dependent constitutive laws, {\it J. Geophys. Res.\/}, {\bf
  120}(9), 6365--6385.

\bibitem[Bizzarri \& Bhat(2012)]{Bizzarri2012}
Bizzarri, A. \& Bhat, H., 2012.
\newblock {\it The Mechanics of Faulting: From Laboratory to Real
  Earthquakes\/}, Research Signpost.

\bibitem[Blanpied et~al.(1998)Blanpied, Marone, Lockner, Byerlee, \&
  King]{blanpied1998}
Blanpied, M., Marone, C., Lockner, D., Byerlee, J., \& King, D., 1998.
\newblock Quantitative measure of the variation in fault rheology due to
  fluid-rock interactions, {\it J. Geophys. Res.\/}, {\bf 103}(B5), 9691--9712.

\bibitem[Bocquet(2013)]{bocquet2013}
Bocquet, L., 2013.
\newblock Friction: An introduction, with emphaisis on some implications in
  winter sports, {\it Sports Physics, edited by C. Clanet (Editions de l'Ecole
  Polytechnique, 2013)\/}.

\bibitem[Brace \& Byerlee(1966)]{brace1966b}
Brace, W.~F. \& Byerlee, J.~D., 1966.
\newblock Stick-slip as a mechanism for earthquakes, {\it Science\/}, {\bf
  153}(3739), 990--992.

\bibitem[Day(1982)]{day1982a}
Day, S.~M., 1982.
\newblock Three-dimensional simulation of spontaneous rupture: the effect of
  nonuniform prestress, {\it Bull. Seism. Soc. Am.\/}, {\bf 72}(6A),
  1881--1902.

\bibitem[den Hartog \& Spiers(2014)]{Hartog2014}
den Hartog, S. A.~M. \& Spiers, C.~J., 2014.
\newblock A microphysical model for fault gouge friction applied to subduction
  megathrusts, {\it Journal of Geophysical Research: Solid Earth\/}, {\bf
  119}(2), 1510--1529, 2013JB010580.

\bibitem[Dieterich \& Kilgore(1994)]{dieterich1994a}
Dieterich, J. \& Kilgore, B., 1994.
\newblock Direct observation of frictional contacts: New insights for
  state-dependent properties, {\it Pure Appl. Geophys.\/}, {\bf 143}(1),
  283--302.

\bibitem[Dieterich(1979{\natexlab{a}})]{dieterich1979a}
Dieterich, J.~H., 1979{\natexlab{a}}.
\newblock Modeling of rock friction: 1. experimental results and constitutive
  equations, {\it J. Geophys. Res.\/}, {\bf 84}(B5), 2161--2168.

\bibitem[Dieterich(1979{\natexlab{b}})]{dieterich1979b}
Dieterich, J.~H., 1979{\natexlab{b}}.
\newblock Modeling of rock friction: 2. simulation of preseismic slip, {\it J.
  Geophys. Res.\/}, {\bf 84}(B5), 2169--2175.

\bibitem[Dragert et~al.(2001)Dragert, Wang, \& James]{Dragert2001}
Dragert, H., Wang, K.~L., \& James, T.~S., 2001.
\newblock A silent slip event on the deeper cascadia subduction interface, {\it
  Science\/}, {\bf 292}(5521), 1525--1528.

\bibitem[Dugdale(1960)]{dugdale1960}
Dugdale, D., 1960.
\newblock Yielding of steel sheets containing slits, {\it J. Mech. Phys.
  Solids\/}, {\bf 8}, 66--75.

\bibitem[Eshelby(1969)]{eshelby1969}
Eshelby, J.~D., 1969.
\newblock The elastic field of a crack extending non-uniformly under general
  anti-plane loading, {\it J. Mech. Phys. Solids\/}, {\bf 17}(3), 177--199.

\bibitem[Faulkner et~al.(2006)Faulkner, Mitchell, Healy, \& Heap]{Faulkner2006}
Faulkner, D.~R., Mitchell, T.~M., Healy, D., \& Heap, M.~J., 2006.
\newblock Slip on 'weak' faults by the rotation of regional stress in the
  fracture damage zone, {\it Nature\/}, {\bf 444}(7121), 922--925.

\bibitem[Froment et~al.(2014)Froment, McGuire, van~der Hilst, Gouedard, Roland,
  Zhang, \& Collins]{Froment2014}
Froment, B., McGuire, J.~J., van~der Hilst, R.~D., Gouedard, P., Roland, E.~C.,
  Zhang, H., \& Collins, J.~A., 2014.
\newblock Imaging along-strike variations inmechanical properties of the gofar
  transform fault, east pacific rise, {\it Journal of Geophysical
  Research-solid Earth\/}, {\bf 119}(9), 7175--7194.

\bibitem[Gratier et~al.(2014)Gratier, Renard, \& Vial]{Gratier2014}
Gratier, J.~P., Renard, F., \& Vial, B., 2014.
\newblock Postseismic pressure solution creep: Evidence and time-dependent
  change from dynamic indenting experiments, {\it Journal of Geophysical
  Research-solid Earth\/}, {\bf 119}(4), 2764--2779.

\bibitem[Hiramatsu et~al.(2005)Hiramatsu, Honma, Saiga, Furumoto, \&
  Ooida]{Hiramatsu2005}
Hiramatsu, Y., Honma, H., Saiga, A., Furumoto, M., \& Ooida, T., 2005.
\newblock Seismological evidence on characteristic time of crack healing in the
  shallow crust, {\it Geophys. Res. Lett.\/}, {\bf 32}(9).

\bibitem[Ida(1972)]{ida1972a}
Ida, Y., 1972.
\newblock Cohesive force across tip of a longitudinal-shear crack and griffiths
  specific surface-energy, {\it J. Geophys. Res.\/}, {\bf 77}, 3796--3805.

\bibitem[Ide et~al.(2007)Ide, Shelly, \& Beroza]{Ide2007}
Ide, S., Shelly, D.~R., \& Beroza, G.~C., 2007.
\newblock Mechanism of deep low frequency earthquakes: Further evidence that
  deep non-volcanic tremor is generated by shear slip on the plate interface,
  {\it Geophysical Research Letters\/}, {\bf 34}.

\bibitem[Ito \& Obara(2006)]{Ito2006}
Ito, Y. \& Obara, K., 2006.
\newblock Very low frequency earthquakes within accretionary prisms are very
  low stress-drop earthquakes, {\it Geophys. Res. Lett.\/}, {\bf 33}(9).

\bibitem[Ito et~al.(2007)Ito, Obara, Shiomi, Sekine, \& Hirose]{Ito2007}
Ito, Y., Obara, K., Shiomi, K., Sekine, S., \& Hirose, H., 2007.
\newblock Slow earthquakes coincident with episodic tremors and slow slip
  events, {\it Science\/}, {\bf 315}(5811), 503--506.

\bibitem[Johnson et~al.(2012)Johnson, Fukuda, \& Segall]{Johnson2012}
Johnson, K.~M., Fukuda, J., \& Segall, P., 2012.
\newblock Challenging the rate-state asperity model: Afterslip following the
  2011 m9 tohoku-oki, japan, earthquake, {\it Geophysical Research Letters\/},
  {\bf 39}(20), L20302.

\bibitem[Kato \& Tullis(2001)]{kato2001}
Kato, N. \& Tullis, T.~E., 2001.
\newblock A composite rate-and state-dependent law for rock friction, {\it
  Geophys. Res. Lett.\/}, {\bf 28}(6), 1103--1106.

\bibitem[Kostrov(1966)]{kostrov1966}
Kostrov, B., 1966.
\newblock Unsteady propagation of longitudinal shear cracks, {\it J. Appl.
  Math. Mech.-USS.\/}, {\bf 30}, 1241--1248.

\bibitem[Kostrov(1964)]{kostrov1964}
Kostrov, B.~V., 1964.
\newblock {Selfsimilar problems of propagation of shear cracks}, {\it J. Appl.
  Math. Mech.-USS.\/}, {\bf 28}(5), 1077--1087.

\bibitem[Lapusta \& Barbot(2012)]{Lapusta2012}
Lapusta, N. \& Barbot, S., 2012.
\newblock Models of earthquakes and aseismic slip based on laboratory-derived
  rate and state friction laws, in {\em The mechanics of faulting: From
  laboratory to earthquakes\/}.

\bibitem[Leonov \& Panasyuk(1959)]{leonov1959}
Leonov, M.~Y. \& Panasyuk, V.~V., 1959.
\newblock Development of the smallest cracks in a solid, {\it Prikl. Mekh.\/},
  {\bf 5}(4), 391--401.

\bibitem[Linker \& Dieterich(1992)]{linker1992}
Linker, M. \& Dieterich, J., 1992.
\newblock Effects of variable normal stress on rock friction: Observations and
  constitutive equations, {\it J. Geophys. Res.\/}, {\bf 97}(B4), 4923--4940.

\bibitem[Marone(1998)]{marone1998}
Marone, C., 1998.
\newblock Laboratory-derived friction laws and their application to seismic
  faulting, {\it Ann. Rev. Earth Planet. Sci.\/}, {\bf 26}(1), 643--696.

\bibitem[Mitchell \& Faulkner(2008)]{Mitchell2008}
Mitchell, T.~M. \& Faulkner, D.~R., 2008.
\newblock Experimental measurements of permeability evolution during triaxial
  compression of initially intact crystalline rocks and implications for fluid
  flow in fault zones, {\it Journal of Geophysical Research-solid Earth\/},
  {\bf 113}(B11), B11412.

\bibitem[Mitchell \& Faulkner(2009)]{Mitchell2009}
Mitchell, T.~M. \& Faulkner, D.~R., 2009.
\newblock The nature and origin of off-fault damage surrounding strike-slip
  fault zones with a wide range of displacements: A field study from the
  atacama fault system, northern chile, {\it Journal of Structural Geology\/},
  {\bf 31}(8), 802--816.

\bibitem[Obara(2002)]{Obara2002}
Obara, K., 2002.
\newblock Nonvolcanic deep tremor associated with subduction in southwest
  japan, {\it Science\/}, {\bf 296}(5573), 1679--1681.

\bibitem[Okubo et~al.(2019)Okubo, Bhat, Rougier, Marty, Schubnel, Lei, Knight,
  \& Klinger]{Okubo2019}
Okubo, K., Bhat, H.~S., Rougier, E., Marty, S., Schubnel, A., Lei, Z., Knight,
  E.~E., \& Klinger, Y., 2019.
\newblock Dynamics, radiation, and overall energy budget of earthquake rupture
  with coseismic off-fault damage, {\it Journal of Geophysical Research: Solid
  Earth\/}, {\bf 124}(11), 11771--11801.

\bibitem[Outerbridge et~al.(2010)Outerbridge, Dixon, Schwartz, Walter, Protti,
  Gonzalez, Biggs, Thorwart, \& Rabbel]{Outerbridge2010}
Outerbridge, K.~C., Dixon, T.~H., Schwartz, S.~Y., Walter, J.~I., Protti, M.,
  Gonzalez, V., Biggs, J., Thorwart, M., \& Rabbel, W., 2010.
\newblock A tremor and slip event on the cocos-caribbean subduction zone as
  measured by a global positioning system (gps) and seismic network on the
  nicoya peninsula, costa rica, {\it Journal of Geophysical Research: Solid
  Earth\/}, {\bf 115}(B10), B10408.

\bibitem[Palmer \& Rice(1973)]{palmer1973}
Palmer, A.~C. \& Rice, J.~R., 1973.
\newblock Growth of slip surfaces in progressive failure of over-consolidated
  clay, {\it Proc. R. Soc. Lond. Ser-A\/}, {\bf 332}, 527--548.

\bibitem[Peng \& Gomberg(2010)]{Peng2010}
Peng, Z. \& Gomberg, J., 2010.
\newblock An integrated perspective of the continuum between earthquakes and
  slow-slip phenomena, {\it Nature Geoscience\/}, {\bf 3}, 599--.

\bibitem[Perrin et~al.(2016)Perrin, Manighetti, Ampuero, Cappa, \&
  Gaudemer]{Perrin2016}
Perrin, C., Manighetti, I., Ampuero, J.-P., Cappa, F., \& Gaudemer, Y., 2016.
\newblock Location of largest earthquake slip and fast rupture controlled by
  along-strike change in fault structural maturity due to fault growth, {\it
  Journal of Geophysical Research: Solid Earth\/}, {\bf 121}(5), 3666--3685,
  2015JB012671.

\bibitem[Perrin et~al.(1995)Perrin, Rice, \& Zheng]{perrin1995}
Perrin, G., Rice, J.~R., \& Zheng, G., 1995.
\newblock Self-healing slip pulse on a frictional surface, {\it J. Mech. Phys.
  Solids\/}, {\bf 43}(9), 1461--1495.

\bibitem[R.~Burridge(1967)]{burridge1967}
R.~Burridge, L.~K., 1967.
\newblock Model and theoretical seismicity, {\it Bulletin of the Seismological
  Society of America\/}, {\bf 57}(3), 341--371.

\bibitem[Rabinowicz(1958)]{rabinowicz1958}
Rabinowicz, E., 1958.
\newblock The intrinsic variables affecting the stick-slip process, {\it Proc.
  Phys. Soc.\/}, {\bf 71}(4), 668.

\bibitem[Rice(1993)]{rice1993}
Rice, J.~R., 1993.
\newblock Spatio-temporal complexity of slip on a fault, {\it J. Geophys.
  Res.\/}, {\bf 98}(B6), 9885--9907.

\bibitem[Rice(2006)]{rice2006}
Rice, J.~R., 2006.
\newblock Heating and weakening of faults during earthquake slip, {\it J.
  Geophys. Res.\/}, {\bf 111}(B05311).

\bibitem[Romanet(2017)]{romanet2017b}
Romanet, P., 2017.
\newblock {\it Fast algorithms to model quasi-dynamic earthquake cycles in
  complex fault networks\/}, Ph.D. thesis, Institut de Physique du Globe de
  Paris.

\bibitem[Rousset et~al.(2016)Rousset, Jolivet, Simons, Lasserre, Riel, Milillo,
  Cakir, \& Renard]{Rousset2016}
Rousset, B., Jolivet, R., Simons, M., Lasserre, C., Riel, B., Milillo, P.,
  Cakir, Z., \& Renard, F., 2016.
\newblock An aseismic slip transient on the north anatolian fault, {\it
  Geophysical Research Letters\/}, {\bf 43}(7), 3254--3262, 2016GL068250.

\bibitem[Rubin \& Ampuero(2005)]{Rubin2005}
Rubin, A.~M. \& Ampuero, J.~P., 2005.
\newblock Earthquake nucleation on (aging) rate and state faults, {\it Journal
  of Geophysical Research-solid Earth\/}, {\bf 110}(B11), B11312.

\bibitem[Ruina(1983)]{ruina1983}
Ruina, A., 1983.
\newblock Slip instability and state variable friction laws, {\it J. Geophys.
  Res.\/}, {\bf 88}(10), 359--370.

\bibitem[Schmitt et~al.(2011)Schmitt, Segall, \& Matsuzawa]{schmitt2011}
Schmitt, S., Segall, P., \& Matsuzawa, T., 2011.
\newblock Shear heating-induced thermal pressurization during earthquake
  nucleation, {\it J. Geophys. Res.\/}, {\bf 116}(B6).

\bibitem[Scholz(1998)]{Scholz1998}
Scholz, C.~H., 1998.
\newblock Earthquakes and friction laws., {\it Nature\/}, {\bf 391}, 37--42.

\bibitem[Segall \& Bradley(2012)]{segall2012b}
Segall, P. \& Bradley, A.~M., 2012.
\newblock The role of thermal pressurization and dilatancy in controlling the
  rate of fault slip, {\it J. Appl. Mech.\/}, {\bf 79}(3), 031013.

\bibitem[Segall \& Rice(1995)]{segall1995}
Segall, P. \& Rice, J.~R., 1995.
\newblock {Dilatancy, compaction, and slip instability of a fluid-infiltrated
  fault}, {\it J. Geophys. Res.\/}, {\bf 100}(B11), 22155--22171.

\bibitem[Sibson(1994)]{Sibson1994}
Sibson, R.~H., 1994.
\newblock Crustal stress, faulting and fluid flow, {\it Geological Society,
  London, Special Publications\/}, {\bf 78}(1), 69--84.

\bibitem[Thomas \& Bhat(2018)]{Thomas2018a}
Thomas, M.~Y. \& Bhat, H.~S., 2018.
\newblock Dynamic evolution of off-fault medium during an earthquake: A
  micromechanics based model, {\it Geophysical Journal International\/}, {\bf
  214(2)}, 1267--1280.

\bibitem[Thomas et~al.(2014{\natexlab{a}})Thomas, Avouac, Champenois, Lee, \&
  Kuo]{Thomas2014b}
Thomas, M.~Y., Avouac, J.-P., Champenois, J., Lee, J.-C., \& Kuo, L.-C.,
  2014{\natexlab{a}}.
\newblock Spatiotemporal evolution of seismic and aseismic slip on the
  longitudinal valley fault, taiwan, {\it Journal of Geophysical Research-solid
  Earth\/}, {\bf 119}, 5114--5139.

\bibitem[Thomas et~al.(2014{\natexlab{b}})Thomas, Avouac, Gratier, \&
  Lee]{Thomas2014c}
Thomas, M.~Y., Avouac, J.-P., Gratier, J.-P., \& Lee, J.-C.,
  2014{\natexlab{b}}.
\newblock Lithological control on the deformation mechanism and the mode of
  fault slip on the longitudinal valley fault, taiwan, {\it Tectonophysics\/},
  {\bf 632}, 48--63.

\bibitem[Thomas et~al.(2017{\natexlab{a}})Thomas, Avouac, \&
  Lapusta]{Thomas2017a}
Thomas, M.~Y., Avouac, J.-P., \& Lapusta, N., 2017{\natexlab{a}}.
\newblock Rate-and-state friction properties of the longitudinal valley fault
  from kinematic and dynamic modeling of seismic and aseismic slip, {\it
  Journal of Geophysical Research-solid Earth\/}, {\bf 122}, 3115--3137.

\bibitem[Thomas et~al.(2017{\natexlab{b}})Thomas, Bhat, \&
  Klinger]{Thomas2017b}
Thomas, M.~Y., Bhat, H.~S., \& Klinger, Y., 2017{\natexlab{b}}.
\newblock Effect of brittle off-fault damage on earthquake rupture dynamics, in
  {\em Fault Zone Dynamic Processes: Evolution of Fault Properties During
  Seismic Rupture\/}, vol. 227, pp. 255--280, eds Thomas, M.~Y., Mitchell,
  T.~M., \& Bhat, H.~S., John Wiley \& Sons, Inc.

\bibitem[Titus et~al.(2006)Titus, DeMets, \& Tikoff]{Titus2006}
Titus, S.~J., DeMets, C., \& Tikoff, B., 2006.
\newblock Thirty-five-year creep rates for the creeping segment of the san
  andreas fault and the effects of the 2004 parkfield earthquake: Constraints
  from alignment arrays, continuous global positioning system, and creepmeters,
  {\it Bulletin of the Seismological Society of America\/}, {\bf 96}(4),
  S250--S268.

\bibitem[Tullis \& Schubert(2015)]{Tullis2015}
Tullis, T.~E. \& Schubert, G., 2015.
\newblock 4.06 - mechanisms for friction of rock at earthquake slip rates, in
  {\em Treatise on Geophysics (Second Edition)\/}, pp. 139--159, Elsevier,
  Oxford.

\bibitem[Walsh(1965{\natexlab{a}})]{Walsh1965}
Walsh, J.~B., 1965{\natexlab{a}}.
\newblock The effect of cracks in rocks on poisson's ratio, {\it J. Geophys.
  Res.\/}, {\bf 70}(20), 5249--5257.

\bibitem[Walsh(1965{\natexlab{b}})]{Walsh1965a}
Walsh, J.~B., 1965{\natexlab{b}}.
\newblock The effect of cracks on the compressibility of rock, {\it J. Geophys.
  Res.\/}, {\bf 70}(2), 381--389.

\bibitem[Wibberley et~al.(2008)Wibberley, Yielding, \& Di~Toro]{Wibberley2008}
Wibberley, C.~A., Yielding, G., \& Di~Toro, G., 2008.
\newblock Recent advances in the understanding of fault zone internal
  structure: A review, {\it Geological Society, London, Special
  Publications\/}, {\bf 299}(1), 5--33.

\bibitem[Zhuravlev(2013)]{zhuravlev2013}
Zhuravlev, V.~P., 2013.
\newblock On the history of the dry friction law, {\it Mechanics of solids\/},
  {\bf 48}(4), 364--369.

\end{thebibliography}
\end{document}